\documentclass[12pt, preprint]{aastex}

\newcommand\chisq{\mbox{$\chi^2$}}
\newcommand\acf{\mbox{$w (\theta)$}}
\newcommand\acfls{\mbox{$w_\mathrm{LS}(\theta)$}}
\newcommand\Aw{\mbox{$A_w$}}
\newcommand\Awm{\mbox{$A_w^\mathrm{mask}$}}
\newcommand\bse{\mbox{$\widehat{\mathrm{se}}(A_w)$}}
\newcommand\thtij{\mbox{$\theta_{ij}$}}
\newcommand\ic{\mbox{$\mathrm{IC}$}}
\newcommand\ca{\mbox{$c_A$}}
\newcommand\syer{\mbox{$\delta_\mathrm{sys}(\Aw)$}}

\def\deg{\ifmmode^\circ\else$^\circ$\fi}

\def\mic{~$\mu$m}

\def\arcs{\ifmmode {''}\else $''$\fi}
\def\arcm{\ifmmode {'}\else $'$\fi}
\def\parcs{\sa=.07em \sb=.03em
     \ifmmode $\rlap{.}$^{\scriptscriptstyle\prime\kern -\sb\prime}$\kern -\sa$
     \else \rlap{.}$^{\scriptscriptstyle\prime\kern -\sb\prime}$\kern -\sa\fi}
\def\parcm{\sa=.08em \sb=.03em
     \ifmmode $\rlap{.}\kern\sa$^{\scriptscriptstyle\prime}$\kern-\sb$
     \else \rlap{.}\kern\sa$^{\scriptscriptstyle\prime}$\kern-\sb\fi}

\def\zphot {\mbox{$z_{phot}$}}

\def\spose#1{\hbox to 0pt{#1\hss}}
\def\simlt{\mathrel{\spose{\lower 3pt\hbox{$\mathchar"218$}}
     \raise 2.0pt\hbox{$\mathchar"13C$}}}
\def\simgt{\mathrel{\spose{\lower 3pt\hbox{$\mathchar"218$}}
     \raise 2.0pt\hbox{$\mathchar"13E$}}}
\def\lsim{\rlap{$<$}{\lower 1.0ex\hbox{$\sim$}}}
\def\gsim{\rlap{$>$}{\lower 1.0ex\hbox{$\sim$}}}

\begin{document}

\title{Wide Field Imaging of the Hubble Deep Field South Region
II: The Evolution of Galaxy Clustering at {$z<1$}\footnotemark}

\footnotetext[1]{Based
on observations obtained at Cerro Tololo Inter-American Observatory,
a division of the National Optical Astronomy Observatories, which
is operated by the Association of Universities for Research in
Astronomy, Inc.\ under cooperative agreement with the National
Science Foundation.}

\author{Harry I. Teplitz\altaffilmark{2},
Robert S. Hill\altaffilmark{3},\\
Eliot M. Malumuth\altaffilmark{3}, 
Nicholas R. Collins\altaffilmark{3},
Jonathan P. Gardner, \\
Povilas Palunas\altaffilmark{4},
Bruce E. Woodgate 
}

\affil{Laboratory for Astronomy and Solar Physics, Code 681, Goddard
Space Flight Center, Greenbelt MD 20771 \\Electronic mail:
hit@binary.gsfc.nasa.gov, }

\altaffiltext{2}{NOAO Research Associate}
\altaffiltext{3}{Raytheon ITSS Corp., Lanham MD 20706}
\altaffiltext{4}{Department of Physics, Catholic University of
America, Washington DC 20064}

\begin{abstract}
  
  We present the galaxy-galaxy angular correlations as a function
  of photometric redshift in a deep-wide galaxy survey centered on
  the Hubble Deep Field South. Images were obtained with the Big
  Throughput Camera on the Blanco 4m telescope at CTIO, of 1/2
  square degree in broad-band $uBVRI$, reaching $\sim 24$~mag.
  Approximately 40,000 galaxies are detected in the survey.  We
  determine photometric redshifts using galaxy template fitting to
  the photometry.  Monte Carlo simulations show that redshifts from
  these data should be reliable out to $z\sim 1$, where the
  4000~\AA~break shifts into the $I$~band. The inferred redshift
  distribution, $n(z)$, shows good agreement with the distribution
  of galaxies measured in the HDF North and the Canada-France
  Redshift Survey.  After assigning galaxies to redshift bins with
  width $\Delta z = 0.33$, we determine the two point angular
  correlation function in each bin.  We find that the amplitude of
  the correlation, $A_w$, drops across the three bins to redshift
  $z\sim 1$. Simple $\epsilon$~models of clustering evolution fit
  this result, with the best agreement for $\epsilon=0$.  Hierarchical
  cold-dark-matter models best fit in a low density,
  $\Lambda$-dominated universe.

\end{abstract}


\section{Introduction}

The Hubble Deep Field (Williams et al.\ 1996; HDFN) presented the
deepest view of the universe in what is now the best studied 5
square arcminutes of the sky. With the addition of the Space
Telescope Imaging Spectrograph (Woodgate et al.\ 1998; STIS) and
the Near Infrared Camera and MultiObject Spectrometer (Thompson et
al. 1998; NICMOS), the opportunity was presented to expand on the
HDF program with an ambitious observation of the field around the
z=2.25 QSO J2233-6033 (Boyle 1997). This Hubble Deep Field South
(Williams et al.\ 2000; HDFS) provided a spectroscopic investigation
of the matter along the line of sight to the QSO (Ferguson et
al.\ 2000), the deepest image ever taken at optical wavelengths
(Gardner et al.\ 2000), and deep optical and IR images of parallel
fields (Casertano et al.\ 2000; Fruchter et al.\ 2000).

In 1997 September, we obtained broad-band $uBVRI$ imaging of half of a
square degree around the HDFS. We have cataloged over 40,000
galaxies\footnote[5]{The data are publically available on the web at
  http://hires.gsfc.nasa.gov/$\sim$research/hdfs-btc/.}. The first
results from the project (Palunas et al.\ 2000a, hereafter Paper I)
were a study of QSOs in this field. Here we present a measurement of
the evolution of large-scale structure in the universe, using the
angular correlations of galaxies as a function of photometric
redshifts determined from our multi-color data. We will also publish
the catalog in Palunas et al.\ (2000b; hereafter Paper III).

In the standard hierarchical model of galaxy formation and evolution,
galaxy clustering traces overdense regions in the distribution of dark
matter.  High-redshift galaxies occur at the most extreme peaks in the
density field and are thus biased tracers of mass---i.e., luminous
matter is more strongly clustered than dark matter (Kaiser 1984).
Subsequently, the clustering of the dark matter increases more rapidly
than that of galaxies, so that the two distributions approach one
another and are more similar today than in the past.  In other words,
the galaxy clustering bias is less.  The way in which the bias
decreases is a function of the cosmological parameters $\Omega,
\Lambda,$~and $P(k)$~(e.g.  Kauffman et al. 1999).  Therefore, by
measuring the evolution of the galaxy-galaxy correlation function, we
can constrain models of structure formation.

Clustering has often been estimated from catalogues of two-dimensional
sky coordinates of galaxies in order to infer the three-dimensional,
physical correlation function, $\xi(r)$, using Limber's equation
(e.g., Peebles 1980).  An important recent study of this type is the
APM Galaxy Survey (Maddox et al. 1996).  On the other hand,
spectroscopic surveys have allowed direct measurement of
$\xi(r)$~for local galaxies over a wide field (e.g. Lin et al.\ 1996).
Studies at high redshift have had to rely on pencil-beam redshift
surveys (e.g. Giavalisco et al.\ 1998) or wide-field photometric
surveys in a single filter (Postman et al.\ 1998).  Photometric
redshifts now make it possible to combine the best of both techniques.
While not accurate enough to determine $\xi(r)$, they are sufficient
to assign the galaxies to redshift bins and study the evolution of
\acf~with redshift directly, rather than inferring the evolution from
varying the magnitude limit (Connolly, Szalay, \& Brunner 1998;
Arnouts et al. 1999; Brunner, Szalay, \& Connolly 2000).  We have used
this method to trace the evolution of clustering to z=1.

In section 2, we summarize the observations and data reduction.  In
section 3 we discuss the techniques used in the photometric redshift
analysis.  In section 4 we present the properties of the estimated
redshifts. In section 5 we discuss the correlation of the cataloged
galaxies as a function of redshift.  We present our conclusions in
section 6. Unless otherwise noted, a cosmology of $H_0=65$ km
s$^{-1}$~Mpc$^{-1}$, $\Omega_M=0.3$, $\Omega_{\Lambda}=0.7$~is
assumed.

\section{Observations}

  We obtained $uBVRI$~imaging using the
Big Throughput Camera (Wittman et al.\ 1998; BTC) on the Blanco
4-m telescope at CTIO.  Observations were taken over three nights,
1998 September 16-18.  Integration times and seeing conditions are
listed in Table \ref{tab:  obs}.

The BTC consists of four 2048$\times$2048 pixel CCDs spaced in a grid,
with a plate scale of 0.43\arcs/pixel.  There are 5.4{\arcm}
wide gaps between the detectors, so it is necessary to observe in
a dithered-grid pattern.  The telescope was pointed at a grid of
positions designed to fill in the gaps while maximizing the exposure
time on a region around the HDFS quasar.  We used 4 main positions
spaced by 11.5\arcmin. At each position, small offsets
were made ($\sim 20\arcs$) in order to measure the sky accurately
in the object frames.  The total area surveyed (to varying
signal-to-noise ratio, SNR) was 2116$\Box$\arcm.

Initial data reduction is carried out using standard IRAF\footnote[6]{IRAF 
is distributed by NOAO, which is operated by AURA Inc., under
contract to the NSF} procedures and {\sc btcproc}, a pipeline
reduction package for BTC data (Wittman et al.\ 1998). The  standard
procedures include the usual CCD processing routines, such as bias
subtraction, flat-fielding, and sky subtraction.  The flat fields
are created as ``super-flats'' from median combinations of all data
taken throughout the night on a given chip in a given filter. 

The large format BTC images have substantial optical distortions which
affect the relative positions of objects, the plate scale, and the
shapes of objects across the field of view. {\sc btcproc}~is designed
to correct these distortions.
First, a smoothed sky frame is subtracted from each exposure (after
the usual flat-fielding, etc.).  Then, preliminary source catalogs
for each image are created in {\sc btcproc} using {\sc SExtractor}
(Bertin \& Arnouts 1996). An initial geometric correction is applied
to each catalog using an $R$-band astrometric solution provided by
the BTC team (Wittman 1999; private communication). The catalog
photometry is also corrected for variations in the plate scale.
All of the preliminary catalogs are then matched to a fiducial position, in our
case to the first $R$-band exposure, and a new astrometric solution is
calculated for each image.  The solution for each image is
parameterized with four two-dimensional, cubic polynomials, one for
each chip.  An additional correction for geometric distortion in
the shapes of objects is calculated to circularize the point spread
function of stars selected from the catalogs. 

The photometry from
the initial catalogs is used to match the relative photometry
between images and chips. Distortion corrected images are registered
and then scaled by the photometry of stars in common.  Finally,
the registered, scaled images are combined by taking the median
of each pixel value.
The combined images are aligned (shifted and rotated) to J2000
coordinates, using astrometric observations of the HDFS field
(Zacharias et al.\ 1998). We measure an RMS positional error of 
0.07 arcseconds (0.16 pixels) in the astrometry of 223 stars.  

A final catalog is constructed, again using {\sc SExtractor}.  In order to
obtain color information through uniform apertures, objects are
identified in each band and then measured on the same pixels in the
other bands.  For objects with strong $R$-band detections, that aperture
is adopted for the photometry of the object; otherwise the highest
SNR band is allowed to define the object.  

The BTC data are observed through the Sloan $u$~filter, the Johnson
$B$~and $V$~filters, and the Cousins $R$~and $I$~filters. We fix our
photometric system to that used by Landolt (1992); that is, Johnson
$UBV$~and Cousins $RI$.  The photometry is calibrated using
observations of the Landolt (1992) standard fields 95 and 113.
Multiple stars are measured on each chip and photometric growth curves
are constructed.  Comparison of the Johnson and Sloan $u$~filters
demonstrates that small color corrections are sufficient to place our
observations onto the same system as Landolt.  Photometric errors are
taken from the {\sc SExtractor} catalogs.  For use in this paper, all
photometric data are converted from counts to flux density,
$f_{\lambda}$, using the photometric zeropoints provided by Bessel
(1990).

Star-galaxy separation is performed using the same method as in Paper
I.  Objects are classified based on their magnitude, FWHM, and {\sc
  SExtractor}'s \emph{class}~parameter.  Objects brighter than
$R=20$~are considered stars if they have $\mathrm{FWHM}<1.8$\arcs~in
the $R$-band.  For objects with $20<R\le 23$, stars are defined as
objects with $\mathrm{FWHM}_R<1.8$\arcs~and \emph{class}$ > 0.95$.  At
faint magnitudes ($R>23$), objects with $\mathrm{FWHM}_R<1.9$\arcs~and
\emph{class}$>0.92$~are classified as stars.  The classification is
limited by the poor seeing of the observations ($FWHM=1.75$~in the
$R$-band); however, the star-galaxy separation is verified by
comparing our classifications to those in the WFPC2 flanking fields
(Lucas et al.  2000).  For objects with $R<24$, we find that the two
data sets agree on the classification of 92\%~of the objects detected.
The number of objects classified as stars is within $\sim 5$\%~of the 
prediction from Galactic star counts (Bahcall \& Soneira 1981).

The catalog contains 53,657 objects with $V<25.0$, of which 46,983 are
classified as galaxies.  For the photometric redshift analysis, we
select the subset of the galaxies which are clearly detected
($3\sigma$) in $BVR$~and either $u$~or $I$, for a total of 29,023
galaxies.

\section{Photometric Redshifts}

Photometric redshifts have been successfully determined for many
catalogs of galaxy photometry, most notably the HDFN observations (see
Koo 1985, Brunner et al.\ 1997, Connolly et al.\ 1997, Lanzetta et
al.\ 1996, Giallongo et al.\ 1998, Arnouts et al.\ 1999; Bolzonella et
al. 2000; for a review see Hogg et al.\ 1998).  The spectral energy
distribution (SED) of a candidate galaxy is compared to a database of
template spectra; the best fit between the two is considered to give
the photometric redshift. The idea, while simple, depends on the
complicated question of choosing the proper template spectra.

Lanzetta et al.\ (1996) have found that ``redshifting'' empirical spectra
of four morphological types of local galaxies provides a sensitive means of
fitting SEDs.  The observed spectra adopted as input to their technique
are those of Coleman, Wu \& Weedman (1980; hereafter
CWW).  Brunner et al.\  (1997) prefer to use galaxy spectra actually
measured at various redshifts as the template database.  The
limitation to that technique is the dearth of $1<z<2$~galaxy spectra
available for comparison.  Finally, many groups (e.g. Giallongo et al.
1998) have used the synthetic spectra of Bruzual \& Charlot's GISSEL96
models (see Bruzual \& Charlot 1993) as templates.  Although the
advantage to synthetic spectra is the inclusion of evolution in the
galaxy spectra, Lanzetta et al.\  argue that evolution mostly has the
effect of moving the SED from one galaxy type to another.

We adopt the simplest of these approaches to fitting the photometric
data.  We compare the measured points to an ``augmented CWW template
set'', constructed from the CWW observed spectra and a single, bluer
GISSEL96 model.  The starburst model is taken from the GISSEL96
library with 0.5 Gyr age, solar metallicity, and a Salpeter IMF.  The
CWW spectra lack a blue starburst spectrum like the ones seen in
(e.g.) the Kinney et al.\ (1993) library.  We expect that the CWW
spectra will better represent evolved, low-redshift galaxies than the
youngest high-redshift ones, which are often strikingly similar to low
redshift starbursts (see Conti, Leitherer \& Vacca 1996).  Thus, we
include CWW spectra for the likely low-z objects and a GISSEL96
starburst model spectrum for the likely high-z ones.  A similar
approach is taken by Gwyn et al.\ (1999), though they include more
than one starburst.  Following the usual approach to template
selection, we do not include any AGN in the SEDs (but see
Hatziminaoglou, Mathes, \& Pello 2000).  The contamination by AGN is
expected to be very small.  In 12\mic-selected samples, which are
significantly less biased against finding Seyferts than our optically
selected sample, only $\sim 7$\%~are spatially resolved AGN (Rush,
Malkan, \& Spinoglio 1993).

We now review the specifics of our technique.  For all the template
spectra, \chisq~is calculated over 102 redshifts. The redshift grid is
evenly spaced in the square root of redshift.  Other authors choose
linearly or logarithmically spaced redshift points; however, we find
that the square root is the best compromise to obtain sufficient
coverage at both large and small redshifts.

The $\chi^2$~test comparing the model SEDs to the observed photometric
data is performed with the usual procedure. For each object observed, 
\begin{displaymath}
\chi^2 = \sum_i \left[ {F_{O,i} - nc\cdot F_{T,i}} 
                       \over
                       {\sigma_{i}} \right]^2 \ ,
\end{displaymath}  
where $F_{O}$~is the observed flux for each filter $i$, with uncertainty 
$\sigma_{i}$, and $F_{T}$~is the
flux predicted by the template model spectrum.  The template fluxes
are normalized to the observed data by a constant, $nc$, which is 
determined (following Giallongo et al.\ 1998) to minimize $\chi^2$, such that
d$\chi^2/$d$nc =0$: 
\begin{displaymath}
nc = \sum_i \left[ {F_{O,i}\cdot F_{T,i}} \over {\sigma^{2}}  \right]
     \Big/
     \sum_i \left[ F_{T,i}^{2} \over {\sigma^{2}}\right]
\end{displaymath}

Given the varying sensitivity between filters, many objects are not detected
in one or more bands.  This problem is particularly evident in the $u$-band
data, given the difficulty of obtaining good $u$-band photometry from the
ground.  In cases where no object is detected in a given filter, we use
the measured flux from the region where the object is detected in other filters.
Since our data are distortion corrected with a large set of astrometric
positions, the registration should prevent this technique from introducing
spuriously high flux from other objects (splitting of faint objects in such
cases is handled by {\sc SExtractor}). 

The hundreds of \chisq~values form a surface, where the independent
variables are the redshift and the galaxy template and the dependent
variable is $\chi_{\nu}^2$, the reduced \chisq~(see e.g. Bevington
1969).  These surfaces typically contain 3-6 minima of varying depth.
Each minimum corresponds to a photometric redshift for the object
being tested.  The most likely inferred luminosity of each galaxy is
favored: rather than simply eliminating the extreme solutions
($L_*<0.001$~or $L_*>10$), we form a weighting function, $P_{LF}$, by
which we divide the $\chi_{\nu}^2$~values.  This function takes the
form of a Schechter function (thus values near $L_*$~are weighted
highly, while the low luminosity end rolls off slowly, and the high
luminosity end drops precipitously --- see Figure \ref{fig: zerr}).
Specifically, $M_{v}^*=-21.5$~and $\alpha=-1.25$.  This
weighting function is not meant to represent the actual luminosity
function of the galaxies, and no luminosity evolution is
included.  Rather $P_{LF}$~is used to discriminate between extremely different
minima in the common sense way that would be applied if each
\chisq-map were inspected by a person.  Cases are rare in which
multiple minima of substantially different redshift fall in the most
probable ($P_{LF} > 0.5$) regime.

\subsection{Calibration of the Photometric Redshift Technique}
\label{sec:  zphot cal}

Accurate redshift estimation requires wavelength coverage of the
strong spectral features.  For example, at low redshifts the strongest
feature is the 4000\AA~break. However, at $z>1.2$~this break redshifts
into the near-IR.  As a result, many surveys (e.g. Connolly et al.\ 
1997) have demonstrated the need for IR data to obtain accurate
$z_{phot}$~for the interesting $1<z<2.5$~era.  At $z\sim 2.5$~the
912\AA~Lyman break redshifts into the blue end of the optical.  The
signal from this feature has led to the successful $U$- and $B$-band
dropout technique (e.g. Steidel et al.\ 1996, Madau et al.\ 1996) of
identifying Lyman break galaxies (hereafter LBGs) and the use of
photometric redshifts to identify galaxies at $z\simgt 5$~(Lanzetta
1999).

In the case of the BTC data presented in this paper, we have  
$uBVRI$~photometry for all galaxies in the catalog.  As a result,
our analysis is limited to galaxies with $z_{phot} \le 1$, where our
spectral coverage is sufficient to detect the characteristic features. 

To calibrate the accuracy of our procedure, we have created catalogs
of simulated observations.  The basis for these simulations is 
the GISSEL96 models at a variety of ages, with solar metallicity.
We expect that the dominant source of error in redshift recovery
will be photometric noise, and so the effect of including non-solar metallicity
input spectra to the simulations would be small.
The GISSEL96 input spectra sample galaxies from E through Sc as well as galaxies
with strong starbursts.  To obtain the simulated catalog, we randomly
sample the distribution of spectroscopically measured redshifts for
bright ($I_{AB}<26$) galaxies in the HDFN and redshift the input spectra.  
This distribution is estimated using a smooth
function fit to the HDFN data which have large gaps at $1<z<2$.
Since our data are not deep enough to detect
$u$-band dropouts, we only model galaxies up to $z=3$.  Simulated
galaxies at $z>2.5$~have Lyman forest blanketing applied using the
decrements of Madau (1995).  The redshifted spectra are scaled to a
random sampling of HDFN $I$-band magnitudes, and random photometric
error is added to each data point, within the expected error bars for our BTC 
data.  Those objects with SNR
sufficient for detection are retained in the simulated catalog.

Figure \ref{fig: zsim residual} shows the residual redshift error
after applying our photometric redshift estimator to each simulated
object.  As expected, the $1<z<2$~regime is the most problematic for
optical data, as the 4000\AA~break redshifts into the near IR, while
the Lyman break is still in the ultraviolet.  We find typical redshift
errors at the 15-20\% level for \zphot$<1$.  However, a number of
$z_{spec}>1$~galaxies are mis-identified as lower redshift objects.
Table \ref{tab: z_mis_id}~shows the proportion of misidentified
objects in several redshift ranges.  This effect would be greatly
reduced by adding IR observations.  In the table, we list
contamination of each bin by objects in bins that are not adjacent;
that is, objects misidentified by the amount of the typical errors are
not considered to be contamination, and some overlap
at the boundaries of the bins is expected.

We also apply our photometric redshift procedure to two catalogs of
spectroscopically confirmed galaxies.  The first is the current set of
HDFN galaxies, as tabulated by Fernandez-Soto et al.\ (1999).  Figure
\ref{fig: z hdfn}~shows the comparison between our \zphot~estimator
applied to the HDFN WFPC2 data and the spectroscopic redshifts. 
Typical redshift residuals are $\Delta z \sim 0.1$~for $z<1$.  The
lower residuals with this data are the result of more precise
photometry.

The second spectroscopic data set is the catalog of AAT spectra for
galaxies around the HDFS (Glazebrook et al.\ 2000; in preparation).
More than 50 galaxies have spectroscopic redshifts.  The objects in
that catalog are in our field, so that a direct comparison with the BTC
photometric redshift results is possible.
Objects that
are not resolved from neighbors or are severely contaminated
from proximity to bright stars are discarded, leaving 31 galaxies.  Of these, 19
have redshifts that Glazebrook et al.\ consider reliable; i.e.,
they have quality flags 3 or 4 which indicate greater than
50\%~probability for the redshift. Figure \ref{fig: z
  glazebrook}~shows the comparison with our inferred redshifts.  
Good agreement is found for most of the AAT sample, since it is restricted to
$z<1.2$.

Also in Figure \ref{fig: z glazebrook}, we compare our photometric
redshift for objects in the HDFS WFPC2 field to estimates made from
the HST data by Yahata et al.\ (2000) and Gwyn et al.\ (2000).
Those groups agree to within $\Delta z \sim 0.2$~on the redshifts of
$\sim 40$~galaxies in the WFPC2 field that we detect and resolve,
and the comparison is made using
those objects. 
Agreement is found within the errors between our estimates and theirs.

\subsection{Errors in Photometric Redshifts}
\label{sec:  zphot error}

For the galaxy catalog, we have calculated the error in the redshift
inferred for individual objects.  These individual errors will not 
be used in the clustering analysis presented in this paper, but may be
of interest in future work.  

One common method of calculating the error on photometric redshift is
to apply the standard $\Delta \chi^2 = 1$~statistic (e.g. Giallongo et
al.\ 1998), i.e., the error, $\delta_z$, is calculated by measuring
the width in redshift of the $\chi^2$~trough at the points where the
value of $\chi ^2_{\nu}=(\chi ^2_{\nu})_{min} + 1$.  This statistic
produces good estimates of the error in $z_{phot}$~as long as the
minimum value of \chisq~is close to unity.

For many galaxies in our catalog, however the value of
\chisq$_{\nu}$~is significantly greater or smaller than one.  Our 
simulations show that this is not an indication of insufficient
resolution in the models;  nor is it the result of systematic
photometric errors (our extensive checks will be presented in Paper III).
Accordingly, we adapt the $\Delta \chi^2 = 1$~method to our data by normalizing our
\chisq~maps such that $(\chi ^2_{\nu})_{min} = 1$~before measuring
the width of the $\Delta \chi^2 = 1$~trough.  For galaxies
having values of \chisq~near unity, this reverts to the usual method.
The values we obtain for $\delta_z$~are consistent with the errors
measured in our simulations (see Table \ref{tab: z_mis_id}~in Section
\ref{sec: zphot cal}).  As expected,  $\delta_z$~roughly increases with 
redshift and with photometric error.

\section{Results}

Figure \ref{fig: zhist}~shows the number-redshift relation, $n(z)$~for
the photometric redshifts of the galaxies in our catalog.  
Our result is compared to the distribution from Yahata et al.\ (2000) for the
HDFS.  The distribution of inferred redshifts agrees fairly well,
although they see a sharp peak at $z\sim 0.5$~that is not present
in our data. This may be the result of the small area over which
their redshifts are estimated.  Glazebrook et al.\  (2000; in 
preparation\footnote[7]{see http://www.aao.gov.au/hdfs/Redshifts/})
see evidence for a cluster at $z=0.585$~in the WFPC2 field.  The 
estimates of $n(z)$~by Yahata et al.  (2000) are for a few hundred
galaxies in the WFPC2 and NICMOS fields, and so could be biased by
such a cluster.  We also compare to the counts for photometric
redshifts in the HDFN (Fernandez-Soto et al.  1999 and the references
therein), and the spectroscopic redshifts from the Canada France
Redshift Survey (Lilly et al.\ 1995).  Our number-redshift counts
show a good agreement with both samples out to $z=1$.

The sample of galaxies for which we can fit redshifts
is characterized by Figure \ref{fig:pzmodel}.  No
brightness cut apart from that implicit in the fitting procedure is
applied.  The top row of plots shows histograms of SED type in each
redshift bin, and the bottom row shows the distribution of $M_R$~for each type.  
The distance modulus of each galaxy is computed using
$\Omega_M=0.3$ and $\Omega_\Lambda=0.7$.  The absolute magnitude is
K-corrected using the best-fit SED.
 Also, a small correction is made for foreground extinction
$E(B-V)=0.027$ (Schlegel, Finkbeiner \& Davis 1998).    If
the SED types are coded as successive integers, then the mean SED
type, which is shown in the top row of plots as a short vertical line,
moves very little between the $z$~ranges. Fernandez-Soto et al.\  (1999)
similarly find little evolution in the median galaxy type, although their
type distribution is centered closer to the irregular galaxy model.

As expected for a magnitude-limited sample, the objects at high
redshift are intrinsically brighter than those at low redshift for a
given SED type.  Thus, any derived quantities as a function of redshift
may be affected by the implicit luminosity selection, and similarly
for different apparent magnitude cuts.  In the case of spatial
correlation statistics, such effects are called luminosity bias (Park
et al. 1994, Kauffmann et al. 1999).

In the highest-redshift bin, the distribution of best-fit SED types is
peaked in the neighborhood of the mean type.  An apparent deficit of
E/S0 galaxies is especially conspicuous.  This effect likely results
from the combination of two factors:  spectral evolution and a selection
bias.  

Spectral evolution affects the distribution of SED types at $z\sim 1$
because our small set of 5 SED templates is static, and each SED is
labeled with the corresponding present-day galaxy type.  Thus a young,
rather blue galaxy at $z \sim 1$ that is fated to become a normal
elliptical might be fit best by an SED appropriate to a present-day
spiral.  Accordingly, we expect that evolution will be reflected in the
choice of template, resulting in a deficit of apparent E/S0 SED types at
higher redshift.  For the HDF-North, using a template set similar to
ours, Fernandez-Soto et al. (1999) find that significantly more galaxies
are matched by the bluer templates at higher redshifts, though they
still do find some galaxies to be classified as E/S0.  However, the HDF
data include the near-infrared $JHK$ passbands, whereas we are limited
to the optical.  

Selection bias may also affect the distribution of SED types.  In
order to fit a photometric redshift, we require a galaxy to be
detected in at least 4 contiguous bands of the set $uBVRI$.  Because
the 4000 \AA~break at $z \sim 1$~is longward of the peak response in
$R$, the rate of inclusion of red galaxies is expected to be less than
that of blue galaxies.  However, the SED type fitted for any galaxy in
this redshift bin is also less meaningful than at lower redshifts
because of relatively large photometric error in most of the bands.
As a result, misclassification may contribute somewhat to the E/S0
deficit.  The redshift itself is not affected either by this ambiguity
or by SED evolution, since it relies primarily on the location of the
4000 \AA~break.

\section{Clustering Properties}
\label{sec:  clustering properties}

\subsection{Definitions}

Our photometric redshift catalog can be used to examine the clustering  
evolution of galaxies at redshifts smaller than unity.  With redshift
uncertainties $\delta_z \sim 0.15$, physical clustering in three
dimensions cannot be evaluated directly; instead, the angular
correlation function is calculated for galaxies selected in redshift 
bins.

One measure of galaxy clustering is the spatial two-point
correlation function $\xi$, which gives the excess probability of
finding a galaxy at a given distance from another galaxy as compared
to a uniformly random location.  The expression for the joint
probability of objects existing in two volume elements $\delta V_i$,
$\delta V_j$ at a separation $r$ is
\begin{displaymath}
\delta P = \mathcal{N}^2[1 + \xi(r)]\delta V_i \delta V_j ,
\end{displaymath}
where $\mathcal{N}$
is the mean space density of objects.  If $\xi=0$, then the objects 
are uniformly distributed.
The clustering of objects on the sky is described by the analogous
joint probability for angular separation,
\begin{displaymath}
\delta P = \mathcal{N}_\Omega^2[1 + \acf]\delta \Omega_i \delta \Omega_j ,
\end{displaymath}
where \acf is angular correlation function, $\delta \Omega_i$, $\delta \Omega_i$ are
elements of solid angle, and $\mathcal{N}_\Omega$ is the mean surface
density of objects.  The angular correlation function is related to
the spatial one by an integration over two lines of sight separated by
angle $\theta$, weighted with a selection function characterizing galaxy
sampling as a function of distance.
This 
integration is described by Limber's equation (Peebles 1980). 

At low redshift and relatively small separations, $\xi$
appears be a power law,
\begin{equation}
\xi(r) = (r/r_0)^{-\gamma}
\label{eq:xi}
\end{equation}
where $r_0$~is the correlation length, and $\gamma$~is the slope.
On scales smaller
than $r=10h^{-1}$Mpc, $\gamma \simeq 1.8$ and $r_0 \simeq 5 h^{-1}$Mpc
(Groth \& Peebles 1977; Davis \& Peebles 1983; Tucker et al.\ 1997).
If $\xi$ is a power law, then so is \acf.  Specifically,
\begin{equation}
\acf = \Aw \theta^{-\delta},  
\label{eq:w}
\end{equation}
where $\delta = \gamma - 1$ (Peebles 1974).

The evolution of galaxy clustering with redshift
may be parameterized as a power of
$(1+z)$:
\begin{displaymath}
\xi(r,z) = (r/r_0)^{-\gamma}(1+z)^{-(3+\epsilon)}.
\end{displaymath}
Clustering is fixed in comoving coordinates if $\epsilon=\gamma-3$, and
it is fixed in proper coordinates if $\epsilon=0$.  On the other hand,
if $\epsilon>0$, then clustering decreases in strength with lookback 
time, as expected from gravitational collapse.  Baugh et al.\  (1999)
roughly estimate
that the prediction of linear perturbation theory, $\epsilon=+0.8$, 
approximately describes the clustering of dark matter, but that
the $\epsilon$~model is not a good description of the clustering of
galaxies.  Angular correlations as a function of redshift can test this 
prediction.  We evaluate \acf~for our data grouped into redshift bins,
and we compare this result to $\epsilon$ models integrated over the same
redshift bins using Limber's equation. 

\subsection{Estimation from Data}

The magnitude of \acf~is a measure of apparent clustering, i.e., of the 
departure of the galaxy sample from a random uniform distribution on the
sky. 

Several algorithms have been developed to estimate \acf~from large data
sets.  We use a common method based on galaxy pair counts.  Let
\thtij~be the angular distance between galaxies $i$ and $j$ in the 
observed sample; then the distribution of all \thtij~normalized to a total
of unity is called the data-data correlation (DD).  Although DD is the
fundamental observed distribution, it is not an estimator because of 
sampling biases.  
Two corrections are required:  one for the geometry of the field
and the other for the relationship between the observed galaxies and the 
edges of the field.  These corrections are included through the 
random-random and data-random correlations (RR and DR, respectively),
which are angular-distance distributions similar to DD.  RR is computed from a
random points distributed uniformly within the usable part of
the observed field, and DR is computed from distances between observed and 
random points.  

In this paper, we apply the estimator suggested by Landy \& 
Szalay (1993),
\begin{displaymath}
\acfls = \frac{\mathrm{DD} - 2\mathrm{DR} + \mathrm{RR}}{\mathrm{RR}}.
\end{displaymath}
This estimator is preferred because its variance is approximately
Poissonian, i.e., near the ideal minimum, and because it is unbiased
in the weak correlation regime (Landy \& Szalay 1993).  This estimator
is one of two that are in wide use currently.  The other is the
Hamilton (1993) estimator, which is symbolically described as
$(\mathrm{DD} \times \mathrm{RR})/\mathrm{DR}^2)-1$.  These estimators
are compared analytically by Landy \& Szalay (1993), and with Monte
Carlo methods by Kerscher, Szapudi, \& Szalay (2000).  Although there
are slight differences between the two estimators, both are clearly
superior to the alternatives.  The Hamilton estimator has a subtle
advantage in the three-dimensional analysis of flux-limited redshift
surveys, whereas the Landy-Szalay estimator is less sensitive to the
number of random points used in the computation.  Therefore, the
Landy-Szalay estimator is preferable for the present
case.

The number of random points used to compute the RR and DR terms is
chosen sufficiently large that the random error in $w_\mathrm{LS}$ is
dominated by the DD term.  The number of random points used to compute
DR is not less than 100 times the number of galaxies in any given run.
The RR term, which describes the geometry of the field irrespective of
the galaxies, is computed separately using 16,000 iterations of 1000
random points apiece.  For all the correlation runs, the field is
masked to omit saturated stars, noise along the edges, and a few other
noisy streaks or blemishes in the interior of the image (Figure
\ref{fig:masks}, panel 0).

Implementation of the algorithm has been verified by processing the
North Ecliptic Pole $K$-band catalogue used by Baugh et al.\ (1996)
and reproducing their plot of \acf.  Verification has also been done
using our data, by comparing \acfls~to the ``ensemble estimator'',
$w_\mathrm{ens}(\theta) = (\langle N_i - N \rangle \langle N_j - N
\rangle)/N^2$, where $N_i$ and $N_j$ are counts in square cells 50
pixels ($21\farcs6$) on a side separated by angle $\theta$, and $N$ is
the mean count over all such cells (Groth \& Peebles 1977, Szapudi \&
Szalay 1998).  The averages are taken over $\theta$~bins just as for
pair-count estimates.  These tests show that the \acfls~algorithm is
implemented correctly.

The value of \acf~is modified by a correction for a bias due to the
finite field size, called the integral constraint (\ic).  For a given
form of \acf, the ratio $\ca=\ic/\Aw$~depends on the field geometry.
Assuming the power-law index $\delta=0.8$, a Monte Carlo integration 
gives $\ca=2.90$ for our field ($\theta$ in degrees).  The correlation
amplitude is calculated as the weighted average over bins of width
$\Delta\log\theta=0.2$ of
$\tilde{A}_w(\theta)=\acfls/(\theta^{-0.8}-\ca)$ for $\log \theta \in
(-3,-0.2)$, equivalent to $\theta \in (3\farcs6, 0\fdg631)$.   Similar
procedures are used by Brainerd, Smail, \& Mould (1995) and Villumsen,
Freudling, \& da Costa (1997). 

Many correlation studies in the literature are based on one or two  
photometric bands, so that redshifts are not available.  In these 
cases, samples are selected by faint magnitude limit, and generally, a 
decrease in correlation strength with the inclusion of more and 
fainter sources is found.  Such samples can be selected from our 
database by ignoring the photometric redshift.  Figure 
\ref{fig:awcompare} shows our results for samples with $19 \le R \le 
23$ and $19 \le R \le 24$, independent both of detection in other 
bands and of the success or failure of the redshift determination.   
The resulting estimates of the correlation strength $\Aw$ are shown  
together with those from several other studies.  Although the values
differ in detail among the various authors, our results are in good 
agreement with the consensus. 

To study the evolution of the clustering strength with time, it is 
preferable to calculate the value of \acf~as a function of redshift. 
To date, surveys relying on spectroscopic redshifts are hindered by 
small sample sizes or limited redshift range, so the approach of using 
photometric redshifts has recently been favored (Brunner et al.\ 1999; 
Arnouts et al.\ 1999).  Adopting this approach, we divide our 
photometric redshift catalog into redshift bins and then examine the 
clustering strength in each bin. The size of the bin is an important 
issue (Arnouts et al.\ 1999).  A bin size of $\Delta z=0.33$~(twice 
the typical error) is adopted here in order to mitigate the 
contamination between adjacent bins.   
 
Figure \ref{fig:wvsth} shows $\log [\acfls + \ic]$ vs. $\log \theta$ for  
three redshift ranges covering $\Delta z=0.33$ between $z$ of $0$ and 
$1$, and for two different values of the faint $R$ cutoff, $23$ and 
$24$.  The error bars are discussed in Appendix \ref{apdx}.  The resulting estimates of  
correlation amplitude, assuming a fixed slope $\delta=+0.8$, are given as a 
function of redshift in Table \ref{tab:twopc} and Figure 
\ref{fig:acfvsz}. The most obvious feature is an apparent decline in 
the correlation amplitude with redshift.  The main focus of our error 
analysis is to assess the significance of this decline.

\subsection{Clustering Evolution \label{sec:evolacf}}

The relativistic version of Limber's equation in the narrow angle 
approximation (Phillips et al.\ 1978) is used to integrate the 
$\epsilon$ models over the redshift selection function $\phi(z)$.  A  
functional form of $\phi$ appropriate for magnitude-limited catalogs 
is given by Efstathiou et al.\ (1991), such that $\phi$ has a single 
parameter, which is the median value of the argument $z$; this $\phi$ 
has the shape roughly of a skewed Gaussian with a high-$z$ tail. 
Villumsen et al.\ (1997) tabulate the median $z$ as a function of 
limiting magnitude $R_\mathrm{max}$; their procedure is followed here. 
In Figure \ref{fig:awcompare} we compare the models with  
values of $A_\mathrm{w}$ calculated form subsets of the BTC galaxies 
detected in $R$ with magnitudes $19 \leq R \leq 23$ and $19 \leq R 
\leq 24$.  The model with $\epsilon=+0.8$ agrees well with these 
results and is in general accord with similar results found in the
literature. 

The integrated $\epsilon$ models depend not only on the normalization 
to the present-day correlation length $r_0$, but also on the adopted
cosmological parameters.  Figure \ref{fig:awcompare} is plotted with
$r_0 = 5.75 h^{-1}$ Mpc, $\Omega_M = 0.3$, and $\Omega_\Lambda=0.7$
(Baugh et al.\ 1999); however, the choice of $r_0 = 4 h^{-1}$ Mpc,
$\Omega_M = 1$, and $
\Omega_\Lambda=0$ (Villumsen et al.\ 1997)
produces a virtually identical plot.  In fact, in using Limber's
equation, $r_0$, $\phi$, and the cosmological parameters can be
manipulated in concert to force a fit with a range of
$\epsilon$~values over a substantial range of limiting magnitude.  The
influence of cosmological parameters on clustering history is explored
by Kauffman et al. (1999).

Figure \ref{fig:acfvsz} shows the correlation amplitude $A_w$ as a
function of the photometric redshift bin.  Again, the 
models are integrated using Limber's equation.  In this case, we 
assume a selection function $\phi$ with a simple form independent of
galaxy brightness, i.e., a tophat of width $\Delta z = 0.33$ convolved
with a Gaussian having $\sigma = 0.2$.  The purpose of $\phi$ is to
model the distribution of actual redshifts included in a given subset
of galaxies.  Here, the tophat is the redshift bin, and the Gaussian
represents the random error in photometric redshift, which broadens
the range of actual redshifts included.  In contrast to Figure
\ref{fig:awcompare}, the $R$-band apparent magnitude limit plays no
role in the models in Figure \ref{fig:acfvsz}, since the selection is made directly
on the basis of observed redshift.  Finer details of the distribution
of actual redshifts within each photometric redshift bin represent
higher-order effects that we do not attempt to model; moreover, the
models intrinsically omit differential clustering biases
such as luminosity bias.

In addition, we plot a galaxy formation model from Baugh et
al.\ (1999) for $R<23.5$ without dust obscuration.  The Baugh et al.
models in general show an upturn in clustering amplitude at
$z\approx1$.  Both Baugh et al.\ (1999) and Arnouts et al.\ (1999)
find observational support for this class of behavior.  Our data do
not extend to high enough redshift to see such an upturn; but at any
rate, they are consistent with the model favored by Baugh et al.\ out
to $z\sim 1$.

The clustering gives an appearance of marginally greater strength for
the brighter galaxies in our measurements, suggestive of luminosity
bias (Park et al.\ 1994; Kauffmann et al. 1999).  This effect is not
seen in the highest redshift bin; however, the errors on the
$R<23$~point are relatively large.  Moreover, the models of Kauffman
et al. (1999) show that luminosity bias itself evolves with
$z$~depending on how the luminosity cut is made.  It is possible that
the bias is less apparent at high redshift where the $R<24$~cutoff is
not faint enough to include the intrinsically fainter (less
clustered) galaxies.

The apparent dearth of E/S0 galaxies in the highest-redshift bin
(Figure \ref{fig:pzmodel} ) suggests that this bin may also be affected by a
morphological clustering bias, such as that seen in the local universe
(Hermit et al.  1996; Santiago \& Strauss 1992; Giovanelli, Haynes, \&
Chincarini 1986; Davis \& Geller 1976).  The local bias is in the
sense that early-type galaxies are more strongly clustered than
late-type galaxies.  Hence, if the highest-redshift bin is biased in
favor of luminous galaxies and against E/S0 galaxies, then a
morphological and a luminosity bias may work in opposite directions.
However, any morphological effect appears to be diluted by its
involving only $\sim20\%$ of the galaxies, if the distribution of
morphologies is extrapolated from that at $z<0.67$.  In any case, the
highest redshift bin is naturally the one most subject to any such
systematic effects and most in need of additional observations.

For the assumed cosmology and local clustering scales, the values of
$\Aw$~that we measure are best fit by $\epsilon=0$~among all the
$\epsilon$~curves, though they are also marginally consistent with the
$\epsilon=0.8$~case.  The measurement for the middle bin
($0.33<z<0.67$) is more than $1\sigma$~below the $\epsilon=0$~model,
while the other data favor that model.  Moreover, the shape of both
$\epsilon \ge 0$~models is similar.  Brunner et al.\ (1999) favor the
$\epsilon=0$~model in a study of clustering vs.  photometric redshift
for 3,000 galaxies over 0.054 square degrees.  Their normalization is
to $r_0=3 h^{-1}$~Mpc, and their conclusion applies equally well to a
universe with $\Omega_M=1$~and to an open universe with
$\Omega_M=0.1$.

Most of the models shown in Figure \ref{fig:acfvsz}~($\epsilon \ge 0$,
Baugh et al.)  are no further than $2\sigma$ away from the data
points, and so cannot be ruled out by these data alone.  By contrast,
Figure \ref{fig:awcompare}, which shows a similar plot for flux-limited samples,
suggests that $\epsilon=+0.8$ is favored over the other models
plotted.  However, a detailed comparison between samples selected by
photometric redshift and those selected solely by flux limit is
probably not meaningful.  The reason is that the clustering strength
of the models is strongly affected by the integration over
redshift (the depth dimension) via the assumed selection function
$\phi(z)$.  In turn, $\phi$ depends on the distribution of intrinsic
galaxy luminosities and colors.  Although we assume a standard form of
$\phi$ for each of our two samples defined solely by flux limit
(Efstathiou et al. 1991), we have no independent check on the
correctness of this assumption.  Similarly, the true selection
function for each of our photometric-redshift bins is also uncertain,
as noted previously.

Our results for the two types of galaxy selection are consistent in
that catalog subsets dominated by the lower redshifts show stronger
clustering than those dominated by the higher redshifts.  The sample
defined solely by $R<23$ (Figure \ref{fig:awcompare}) is more clustered than that
defined solely by $R<24$, and the sample defined by $z_\mathrm{phot} <
0.33$ is more clustered than those with $z_\mathrm{phot}>0.33$.
Furthermore, the selection functions adopted for integrating the
clustering evolution models are at least roughly consistent with one
another, since reasonable models yield correlation strengths in the
range delineated by the observations for both types of catalog
subsample.

Nevertheless, the primary purpose of fitting photometric redshifts is
to be able to model $\phi(z)$ for a given subsample in a way that is
both straightforward and independent of intrinsic galaxy properties.
If the random error in photometric redshift is well characterized,
then the actual selection function of a given sample should be better
determined with this method than with a simple apparent magnitude
cutoff.  Accordingly, our expectation is that the model comparison in
Figure \ref{fig:acfvsz}~is more robust than that in Figure
\ref{fig:awcompare}, to within the error bars.

The ability of studies like this one to discriminate between models of
clustering evolution can be improved by increasing either the
photometric depth or the sky area.  Greater depth would provide a
check for the present measurements both through improved photometry
and through an extension of the redshift range.  A greater sky area
would improve the precision of the clustering amplitude by increasing
the number of galaxy pairs at every redshift.

\section{Conclusions}
\label{sec:  conclusions}

Our data are consistent with a popular scenario for the evolution of
clustering with redshift.  As we noted above, more data will help
distinguish between these cases.

Kauffmann et al.\ (1999) discuss the predictions of hierarchical CDM
models on the observed clustering of galaxies.  They explain that in a
low density, $\Lambda$-dominated universe ($\Lambda$CDM), there is a
``dip'' (their phrasing) in the amplitude of galaxy clustering around
$z\sim 1$.  This dip occurs between the era of strong dark matter
clustering traced by unbiased luminous galaxies (low redshift) and the
era in which the observable galaxies are highly biased tracers of the
densest (earliest clustered) regions.  In a high density universe, the
dip is absent, since structures form later and galaxies are more
highly biased at higher redshifts; in that case, the clustering
strength is constant for $z<1$~and then rises sharply with
redshift.  CDM models also predict that the dip is weaker for
redder, more luminous galaxies.  Arnouts et al.\ (1999) demonstrate
similar features in a wider array of CDM models, but with the same
distinction between high density and low density universes.  In our
data, we find evidence for a decline out to $z\sim 1$, though the lack
of higher redshift points prevents us from testing with our data alone whether 
this is a dip or an indication of a continuing decline.
However, a reliable data point at $z\sim 3$~from the Lyman break
galaxies (Giavalisco et al.\ 1998) suggests that a dip must occur
somewhere in the intervening redshift range.  The strength of this
effect is magnified by the fact that at higher redshifts we appear to
be seeing blue galaxies, for which the dip is expected to be
stronger.

Our data are consistent with $\Lambda$CDM predictions for a low density
universe.  Although the strength of the evolution is debated, this scenario 
has been suggested by several studies (e.g. Postman et al.\ 1998, Le
Fevre et al.\ 1996).  Using the photometric redshift technique, Arnouts
et al.\ (1999) find this kind of clustering evolution in the HDF (but
note that Connolly et al.\  1998 do not).  

These results should be taken as a first look at the power of
using the photometric redshift technique to measure clustering
evolution on a sample of tens of thousands of galaxies on large
scales.  With the new instrumentation now becoming available on medium
and large telescopes, this experiment will be repeated with higher
accuracy out to larger redshifts.  In particular, substantial improvements
will occur when it is practical to obtain deep near-IR data over
similar area.

\acknowledgements

We thank Hsiao-Wen Chen for a sanity check of the photometric
calibration.  This work was supported by the STIS Investigation
Definition Team through the National Optical Astronomical Observatories
and by the Goddard Space Flight Center.

\appendix
\section{Errors in the measurement of the angular correlation function}
\label{apdx}

The sum in quadrature of random and systematic error in $\Aw$ is shown
by the thin error bars in Figure \ref{fig:acfvsz}, and the random
error alone is shown by the thick error bars.  By random error, we mean
that arising from the galaxy counting statistics.  Errors in $\acf$ are
Poissonian in the number of galaxy pairs in each $\theta$ bin (Peebles  
1980, Landy \& Szalay 1993); however, because $\theta$~bins are correlated, 
the naive least-squares error in the weighted average of $\tilde{A}_w(\theta)$
is incorrect.  Rather than trying to find the covariance matrix of all
the $\theta$ bins, a typical approach is to use a resampling method.  

In general, resampling is similar to Monte Carlo methods, in that error
is computed using artificial data sets drawn by chance; however, the
data points are taken from the actual observations, rather than from an
idealized model distribution.  These samples are processed in the same
way as the original data, and the standard deviation among all the
answers is within a scale factor of the random error for the full
observed data set.

A popular resampling method is the bootstrap (Efron \& Tibshirani 
1993; Barrow, Bhavsar \& Sonoda 1984), in which each artificial data 
set is the same size as the observed data set.  Necessarily, some data   
points are duplicated, while others are omitted.  This process is the   
same as drawing the points from a pool consisting of an infinite 
number of replicas of the observed data (Diaconis \& Efron 1983). 
However, its application to $\acf$ is intuitively unsatisfying, 
because the duplicate galaxies are perfectly correlated.  An analysis 
of the bootstrap for two-point correlation is given by Snethlage 
(1999), who finds that the resulting variances are too large. 

Accordingly, we adopt an alternative resampling plan, based on the
drawing of subsamples smaller than the original data set, without
replacement; this method is known technically as the \emph{delete-$d$
jackknife}, where $d$ is the number of points omitted to form each
subsample (Efron \& Tibshirani 1993).  The jackknife is known in general to be
applicable to a wider range of problems than the bootstrap (Davison \&
Hinkley 1997).  Our argument for its use here is that each subsample
represents a physically plausible distribution on the sky that is
similar to the original, except sparser.  The standard deviation of
$\Aw$ fitted from multiple subsamples of a given size is scaled to the
full number of data points, $N$, by multiplying by $N^{-0.5}$.  In this
case, $N \approx N_\mathrm{gal}^2$ is the number of galaxy \emph{pairs},
so that the error is scaled by $N_\mathrm{gal}^{-1}$.  

The results of the delete-$d$ jackknife are shown in Figure 
\ref{fig:jack} for the 6 catalogue samples selected on $R$ and $z$.
Each closed circle represents 50 or 100 jackknife subsamples
of the given size.  The abscissa is the number of galaxies in each
subsample, and the ordinate is the standard deviation of $\Aw$ computed
only from the subsamples of that size.  The line is an unweighted fit
assuming the $N_\mathrm{gal}^{-1}$ scaling, and the open circle is the
extrapolated error for the full number of galaxies.  The closed
triangles give upper and lower limits on the error in $\Aw$; the upper
limit is the simple sum of the errors arising from individual $\theta$~bins,
whereas the lower limit is the sum in quadrature.  These limits
represent the extreme cases of complete correlation (upper) and no
correlation (lower) between $\theta$ bins.  We adopt the extrapolated
jackknife errors as the standard error in correlation amplitude, $\bse$.
These errors are $\sim 2$ times the respective lower limits (similar
to the rule of thumb adopted by  Baugh
et al.\ 1996). For each of the two large samples selected on $R$~magnitude
only, the random error is extrapolated from one subsample containing 15\%~of
the galaxies for $R<24$~and 25\%~for $R<23$.

Systematic as well as random error is assessed.  One
distinction between the two types of error is that the magnitude of
systematic error can only be explored and not thoroughly characterized,
whereas random error is a function of some reasonably well-understood
distribution.  Assessing systematic error involves anticipating possible
problems that are not seen clearly in the measurements; for if the
effects were well-defined, they could be corrected and relegated to the
category of calibration. 

Like random error, systematic error is estimated through a repeated
computation of the correlation strength; however, in this case, the
subsets of the original data are chosen deliberately rather than
randomly.  We take systematic error to be the result of additional
structure imposed on the observed galaxy distribution by photometric
non-uniformity.  Two mechanisms are considered: (1) the dithering and
the mosaicking of data from several chips, which result in variation by
a factor of a few in effective exposure over the image; and (2) the
possibility of insufficient masking around the 2 brightest stars,
resulting in broad, low-amplitude clusters of noise spikes.  

The subsets are chosen by masking out parts of the image.  Figure
\ref{fig:masks} shows thumbnail images of the 9 masks used.  Mask~0 is
the standard $R$-band mask used to omit bright stars and edge effects.
Mask~1 omits a range of columns in which is seen a modest concentration
of galaxies that may be aligned with chip boundaries;  masks $2-4$
include only low, middle, and high ranges of total exposure,
respectively; masks~$5-7$ are similar to $2-4$, with extended masking
around bright stars; and mask~$8$ includes the low and high ranges of
total exposure, while omitting the middle range.  The resulting values
of $\Awm$ for each masked sample are plotted in Figure \ref{fig:syserr}.
The systematic error is computed for each range of $z$ and $R$ as
$\syer = \sum N_m|\Awm-\Aw|/\sum N_m$, where $N_m$ is the 
number of galaxies included by each mask, and $\Aw$ has the same meaning
as above, i.e., the adopted correlation strength as computed from the
full sample.  These errors are given in Table \ref{tab:twopc} and shown
as a component of the error bars plotted in Figure \ref{fig:acfvsz}.  

If two outliers with small image area are ignored, the middle and high 
redshift bins are affected little by systematic error as defined 
above.  However, the low-redshift bin shows a greater variation in 
correlation strength depending on field mask.  This variation has a 
pattern: masks that mostly exclude a vertical stripe in the left half 
of the image, seen in panel 1 of Figure \ref{fig:masks}, give a 
relatively low correlation strength (panels 1,2,4,5,7, and 8 of Figure 
\ref{fig:masks}).  The masks that mostly include this stripe give a 
consistent and relatively high correlation strength (panels 0, 3, and 
6 of Figure \ref{fig:masks}).  This region is selected by eye for 
special treatment in the error analysis because of a very subtle 
enhancement in galaxy surface density seen in certain visual displays, 
and because of a suspicion that this feature might be aligned with CCD 
chip boundaries in the final stacked image.  However, the feature may 
also represent, in part, the real clustering signal that is sought. 
If the density enhancement is real, it is still too 
vaguely perceived to be analyzed as a distinct cluster.

\references

\reference{} Arnouts, S., Cristiani, S., Moscardini, L., Matarrese,
S., Lucchin, F., Fontana, A. \& Giallongo, E. 1999, MNRAS, 310, 540

\reference{} Bahcall \& Soneira 1981, ApJS, 47, 357

\reference{} Barrow, J. D., Bhavsar, S. P., \& Sonoda, D. H. 1984,
MNRAS, 210, 19P

\reference{} Baugh, C. M., Gardner, J. P., Frenk, C. S., \& Sharples,
R. M. 1996, MNRAS, 283, 15

\reference{} Baugh, C. M, Benson, A. J., Cole, S., Frenk, C. S., \&
Lacey, C. G. 1999, MNRAS, 305, L21

\reference{} Bertin, E. \& Arnouts, S. 1996, \aaps, 117, 393

\reference{} Bessell, M. S. 1990, PASP, 102, 1181

\reference{} Bevinton, P. R. 1969, {\it Data Reduction and Error
Analysis for the Physical Sciences}, McGraw-Hill Book Company, New
York, New York

\reference{} Bolzonella, M., Miralles, J.-M., \& Pello, R. 2000, A\&A, in press, astro-ph/0003380

\reference{} Boyle, B. 1997, AAONw, 83, 4

\reference{} Brainerd, T. G., Smail, I., \& Mould, J. 1995, MNRAS, 275, 781

\reference{} Brunner, R. J., Connolly, A. J., Szalay, A. S. \&
Bershady, M. A.,1997, ApJ, 482, L21

\reference{} Brunner, R. J., Connolly, A. J., \& Szalay, A. S. 1999, in
"Photometric Redshifts and High Redshift Galaxies", eds.  R. Weymann,
L. Storrie-Lombardi, M. Sawicki \& R. Brunner, (San Francisco: ASP
Conference Series)

\reference{} Brunner, R. J., Szalay, A. S.,  \& Connolly, A. J. 2000, ApJ, 541, 527

\reference{} Bruzual, A. G. \& Charlot, S.,  1993, ApJ, 405, 538

\reference{} Casertano, S., et al., 2000, AJ, in preparation

\reference{} Coleman, G. D., Wu, C. C., \& Weedman, D. W. 1980, \apjs, 43, 393

\reference{} Connolly, A. J., Szalay, A. S., Dickinson, M., Subbarao,
M. U., \& Brunner, R. J. 1997, ApJ, 486, L11

\reference{} Connolly, A. J., Szalay, A. S., \& Brunner, R.J. 1998,
ApJ, 499, L125

\reference{} Conti, P. S., Leitherer, C., \& Vacca, W. D. 1996, ApJ,
461, L87

\reference{} Davis, M., \& Geller, M. 1976, ApJ, 208, 13

\reference{} Davis, M., \& Peebles, P. J. E. 1983, ApJ, 267, 465

\reference{} Davison, A. C. \& Hinkley, D. V. 1997, Bootstrap methods
and their application (Cambridge: CUP)

\reference{} Diaconis, P. \& Efron, B. 1983, Scientific American, 248, 116

\reference{} Efron, B. \& Tibshirani, R. 1993, An introduction to the
bootstrap (New York: Chapman \& Hall)

\reference{} Efstathiou, G., Bernstein, G., Tyson, J. A., Katz, N., \&
 Guhathakurta, P. 1991, ApJ, 380, L47

\reference{} Ferguson, H. C. et al.\ 2000, AJ, in preparation

\reference{} Fernandez-Soto, A., Lanzetta, K. M., \& Yahil, A. 1999,
ApJ, 513, 34

\reference{} Fruchter, A. S., et al.\ 2000, AJ, in preparation

\reference{} Gardner, J. P, et al.\ 2000, AJ, 119, 486

\reference{} Giallongo, E., D'Odorico, S., Fontana, A., Cristiani, S.,
Egami, E., Hu, E.,\& MCMahon, R. G. 1998, AJ, 115, 2169

\reference{} Giavalisco, M., Steidel, C. C., Adelberger, K. L.,
Dickinson, M. E., Pettini, M., \& Kellogg, M. 1998, ApJ, 503, 543

\reference{} Giovanelli, R., Haynes, M. P., \& Chincarini, G. 1986, ApJ, 300, 77

\reference{} Glazebrook, K. et al.\ 2000, in preparation

\reference{} Groth, E. J. \& Peebles, P. J. E. 1977, ApJ, 217, 385

\reference{} Gwyn, S. D. J. 1999, in ``Photometric Redshifts and High
Redshift Galaxies'', eds. R. Weymann, L. Storrie-Lombardi, M. Sawicki
\& R. Brunner, (San Francisco: ASP Conference Series)

\reference{} Gwyn, S. D. J., et al.\ 2000, in preparation

\reference{} Hatziminaoglou, E., Mathez, G., \& Pello, R. 2000, A\&A, 359, 9

\reference{} Hermit, S., Santiago, B. X., Lahav, O., Strauss, M. A., Davis, M.,
Dressler, A., \& Huchra, J. P. 1996, MNRAS, 283, 709

\reference{} Hogg, D. W. et al.\ 1998, AJ, 115, 1418

\reference{} Kaiser, N. 1984, ApJ, 284, 9

\reference{} Kauffmann, G., Colberg, J. M., Diaferio, A., \& White,
S. D. M. 1999, MNRAS, 307, 529

\reference{} Kerscher, M., Szapudi, I., \& Szalay, A. S. 2000, ApJL, 535, L13

\reference{} Kinney, A. L., Bohlin, R. C., Calzetti, D., Panagia, N., \&
Wyse, R. F. G. 1993, ApJS, 86, 5

\reference{} Koo, D. C. 1985, AJ, 90, 418

\reference{} Landolt, A. U. 1992, \aj, 104, 340

\reference{} Landy, S. D. \& Szalay, A. S. 1993, ApJ, 412, 64

\reference{} Lanzetta, K. M., Yahil, A., \& Fernandez-Soto, A. 1996, Nature, 381, 759

\reference{} Lanzetta, K. M. 1999, BAAS, 195.1905

\reference{} Le Fevre, O, Hudon, D., Lilly, S.J., Crampton, D. Hammer,
F., \& Tresse, L. 1996, ApJ, 461, 534

\reference{} Lilly, S. J., Le Fevre, O., Crampton, D., Hammer, F., \&
Tresse, L. 1995, ApJ, 455, 50

\reference{} Lin, H., Kirshner, R. P., Shectman, S. A., Landy, D.  S.,
Oemler, A., Tucker, D. L., \& Schechter, P. L. 1996, ApJ, 471, 617

\reference{} Lucas, R. A. et al.\ 2000, AJ in preparation

\reference{} Madau, P. 1995, ApJ, 441, 18

\reference{} Madau, P., Ferguson, H. C., Dickinson, M. E., Giavalisco,
M., Steidel, C. C., \& Fruchter, A. 1996, \mnras, 283, 1388

\reference{} Palunas, P., Collins, N. R., Gardner, J. P., Hill, R.
S., Malumuth, E. M., Teplitz, H. I., Smette, A., Williger, G. M.,
Woodgate, B. E., \& Heap, S. R. 2000a, ApJ, in press (Paper I)

\reference{} Palunas, P., et al.\ 2000b, in preparation (Paper III)

\reference{} Park, C., Vogeley, M. S., Geller, M. J., \& Huchra 1994,
ApJ, 569, 585

\reference{} Peebles, P. J. E. 1974, A\&A, 32, 197

\reference{} Peebles, P. J. E. 1980, The large-scale structure of the
Universe.  Princeton University Press, Princeton

\reference{} Phillips, S., Fong, R., Fall, R. S., Ellis S. M., \&
MacGillivray, H. T. 1978, MNRAS, 182, 673

\reference{} Postman, M., Lauer, T. R., Szapudi, I., \& Oegerle, W.  1998, ApJ, 506, 33

\reference{} Rush, B., Malkan, M.A., \& Spinoglio, L. 1993, ApJS, 89, 1

\reference{} Sandage, A. 1997, PASP, 109, 1193

\reference{} Santiago, B. X. \& Strauss, M. A. 1992, ApJ, 387, 9

\reference{} Schlegel, D. J., Finkbeiner, D. P, \& Davis, M. 1998,
ApJ, 500, 525

\reference{} Snethlage, M. 1999, Metrika, 49, 245

\reference{} Steidel, C. C., Giavalisco, M., Pettini, M., Dickinson,
M., \& Adelberger, K. L. 1996, ApJ, 462, L17

\reference{} Szapudi, S. \& Szalay, A. S. 1998, ApJLetters, 494, 41

\reference{} Thompson, R. I., Rieke, M., Schneider, G., Hines, D. C., \&
Corbin, M. R. 1998, ApJ, 492, L95

\reference{} Tucker, D. L., et al. 1997, MNRAS, 285, 5

\reference{} Villumsen, J. V., Freudling W., da Costa, L. N. 1997, ApJ, 481, 578

\reference{} Wittman, D. M., Tyson, J. A., Bernstein, G. M., Lee, R. W.,
Dell'Antonio, I. P., Fischer, P., Smith, D. R., \& Blouke, N. M. 1998,
Proc. SPIE, 3355, 626

\reference{} Williams, R. E., et al.\ 1996, AJ, 112, 1335

\reference{} Williams, R. E., et al.\ 2000, AJ, in preparation

\reference{} Woodgate, B. E., et al.\ 1998, PASP 110, 1183

\reference{} Yahata, N., Lanzetta, K. M, Chen, H.-W., Fernandez-Soto, A., Pascarelle, S. M., Yahil, A., \&
Puetter, R. C. 2000, ApJ in press, (astro-ph/0003310)

\reference{} Zacharias, N., Corbin, T., Zacharias, M., Rafferty, T.,
Seidelmann, P. K., \& Gauss, F. S. 1998, BAAS, 193.7509

\clearpage

\renewcommand{\arraystretch}{.5}
\begin{deluxetable}{llll}

\tablecolumns{4}
 
\tablecaption{Observations}
\tablehead{
\colhead{filter} &
\colhead{exptime/frame} &
\colhead{no. frames} &
\colhead{seeing FWHM (\arcs)} 
}

\startdata

$u$  &     1200  &   13   & 1.90 \nl
B    &      600  &   12   & 2.16 \nl
V    &      300  &   15   & 1.85 \nl
R    &      300  &   16   & 1.75 \nl
I    &      300  &   17   & 1.55 \nl

\enddata

\label {tab:  obs}
\end{deluxetable}
\renewcommand{\arraystretch}{1.0}

\clearpage

\renewcommand{\arraystretch}{.5}
\begin{deluxetable}{lllc}

\tablecolumns{4}
 
\tablecaption{Redshift Error Estimates from Simulations}
\tablehead{
\colhead{$z_{phot}$~range} &
\colhead{mode $z_s - z_p$} &
\colhead{$\sigma _z$} &
\colhead{\% contamination\tablenotemark{a}} 
}

\startdata
0--0.33     &  -0.01 &  0.12 &  3   \nl
0.33--0.67  &   0.01 &  0.16 &  3  \nl
0.67--1.00  &  -0.03 &  0.21 &  8 \nl 
1.00--1.33  &  0.15  &  0.35 &  7  \nl
1.33--2.0   &  -0.41 &  0.58 &  40  \nl
2-3         &   1.1  &  0.72 &  68 \nl 

\enddata \tablenotetext{a}{objects in $z_{phot}$~bin whose true redshift belongs
two or more bins away, for bins with $\Delta z = 1/3$;  for example in the $0.33<z<0.67$~bin, 
nothing at $z<1$~is contamination, but a $z_{true}=1.2$~galaxy would be.}

\label {tab:  z_mis_id}
\end{deluxetable}
\renewcommand{\arraystretch}{1.0}

\clearpage

\begin{deluxetable}{cccccc}

\tablewidth{5in}
\tablecolumns{6}
 
\tablecaption{Two-Point Correlation Results \label {tab:twopc}}
\tablehead{
\colhead{Mag Limits}&
\colhead{$z$ Range}&
\colhead{$N_\mathrm{gal}$}&
\colhead{$\Aw$}&
\colhead{$\bse$}&
\colhead{$\syer$}
}
\startdata
$19\leq R\leq23$ & $0-0.33$    & 1423  & 5.45(-3) & 6.8(-4) & 1.9(-3) \\
                 & $0.33-0.67$ & 3992  & 2.92(-3) & 2.0(-4) & 3.6(-4) \\
                 & $0.67-1$    & 1842  & 1.96(-3) & 3.4(-4) & 6.4(-4) \\
$19\leq R\leq24$ & $0-0.33$    & 2608  & 3.88(-3) & 4.0(-4) & 9.9(-4) \\
                 & $0.33-0.67$ & 6338  & 1.95(-3) & 1.4(-4) & 2.1(-4) \\
                 & $0.67-1$    & 3411  & 1.92(-3) & 2.3(-4) & 2.2(-4) \\
$19\leq R\leq23$ & all         & 9467  & 1.37(-3) & 1.0(-4) & \nodata \\
$19\leq R\leq24$ & all         & 22702 & 6.08(-4) & 3.7(-5) & \nodata \\

\enddata
\end{deluxetable}

\clearpage

\clearpage

\figcaption[]{Values of \chisq~as a function of
  redshift for a typical galaxy (solid line).  Each of the five galaxy
  templates produces a similar plot.  For this galaxy, the late type
  spiral template (CWW) has the lowest value of \chisq, and thus is
  shown here.  The dotted line shows the redshift probability
  function, $P_{LF}$~(normalized to unity at the maximum probability)
  for this V=23.4 galaxy. In this case, notice that it would clearly
  favor the lower redshift trough even if its \chisq~were slightly
  higher.  The vertical dashed line shows the location of the
  \chisq~minimum, $z=0.48$, and the horizontal dashed line shows the
  redshift error calculated by the width of the trough at the point at
  which \chisq~is double the minimum.
  \label{fig: zerr}}

\figcaption[]{The error $\delta _z = z_{true} - z_{phot}$~in the
  photometric redshifts estimates for our simulated catalog of
  galaxies. \label{fig: zsim residual}}

\figcaption[]{ Our photometric redshifts, $z_{phot}$, are plotted
  versus spectroscopic redshifts, $z_{spec}$, for ~ 100 galaxies in
  the HDF-north field.  Our photometric redshifts are based on
  $UBVIJHK$~photometry.  The error bars shown were derived using a
  jackknife technique as described in Section \ref{sec: zphot error}.
  The dotted lines show a 0.2 deviation in the values.
  \label{fig: z hdfn}}

\figcaption[]{$z_{true}$~vs. $z_{phot}$~for galaxies in the HDFS.
  Filled circles represent galaxies for which the spectroscopic
  redshifts are available from the AAT.  Open diamonds represent
  galaxies without spectroscopic redshifts, but for which Gwynn et al.
  (1999) and Yahata et al.\ (2000)  agree on photometric redshift within
  0.2 in redshift.  In both cases, only objects well separated from
  neighbors in our images are considered.  Errors are calculated from
  the \chisq~distribution; see section \ref{sec: zphot error}.  
  \label{fig:  z glazebrook}}

\figcaption[]{The number-redshift relation, $n(z)$, for
  galaxies in our photometric redshift sample.  Redshifts are counted
  in bins of width $\Delta z = 0.2$.  $n(z)$~is also
  plotted for other photometric and spectroscopic redshift samples:
  the HDFN photometric redshifts (Fernandez-Soto et al.\  1999); the
  spectroscopic redshifts of the CFRS; the photometric redshift for
  the HDFS of Yahata et al.\ (2000).  Individual histograms are scaled
  to our $n(z)$~by the total number of objects in the sample.
\label{fig:  zhist}
}

\figcaption{
Galaxy types as a function of redshift bin.  The types are the best-fit
SEDs from the photometric redshift computation, identified respectively
with early-type galaxies (E/SO), spirals (Sbc, Scd), irregulars (Irr),
and starbursts (StarB).  Top panels:  histograms of galaxy type in each
photometric redshift bin; vertical line is mean
galaxy type if types are coded as successive integers.  Bottom panels:
absolute $R$-band magnitude $M_R$ (with K and evolutionary corrections);
each galaxy is plotted as a dot; the box shows the quartile and median
values of $M_R$ for each SED type and redshift range.
\label{fig:pzmodel}}

\figcaption{ Correlation strength vs. magnitude for $R$-magnitude
  limited galaxy samples.  Solid circles: galaxy samples from this
  work with $19 \leq R \leq 23$ and $19 \leq R \leq 24$; other symbols
  represent results from literature, extrapolated if necessary
  assuming that $ w \propto \theta^{-0.8}$.  Abscissa: $R$-band faint
  limit of apparent galaxy magnitude.  Ordinate: correlation strength
  $A_w = w(1\arcdeg)$.  Random error bars are smaller than the plotted
  points.  Lines are models integrated using Limber's equation with
  analytic selection function $\phi$ adopted from Villumsen et al.
  (1997) and assuming $\Omega_M = 0.3$, $\Omega_{\Lambda} = 0.7$, and
  $r_0 = 5.75 h^{-1}$ Mpc.  Dashed: $\epsilon = 1.6$; solid: $\epsilon
  = 0.8$; dotted: $\epsilon = 0$; dash-dot: Baugh et al. (1999)
  semi-analytic model for $R<23.5$ and no dust.
\label{fig:awcompare}}

\figcaption{
Angular correlation functions for BTC-HDFS galaxies selected by apparent
magnitude and redshift.  Abscissa:  log of angular separation $\theta$
in degrees.  Ordinate:  log of correlation function $w(\theta) +
\mathrm{IC}$, where IC is the integral constraint.  Legends give
magnitude and redshift ranges.  Solid lines:  best-fit functions of form
$w = A_w \theta^{-0.8}$ (cf. Table \ref{tab:twopc}); that is, the slope is
fixed and only the fit is only for the amplitude.
\label{fig:wvsth}}

\figcaption{
Correlation strength vs. redshift for samples with $19 \leq R \leq 23$
(squares) and $19 \leq R \leq 24$ (triangles).  Abscissa:  midpoint of
redshift bin.  Ordinate:  correlation strength $A_w = w(1\arcdeg)$,
assuming that $ w \propto \theta^{-0.8}$.  Thick error bars are random
error arising from counting statistics of galaxies.  Thin error bars are
sum in quadrature of random error and an estimate of systematic error
resulting from variation in sensitivity over $R$-band image.  Lines are
models of galaxy evolution for $\Omega_M = 0.3$, $\Omega_{\Lambda} = 0.7$,
and $r_0 = 5.75 h^{-1}$ Mpc.  Models are integrated using Limber's
equation; the selection function $\phi$ is a tophat of width $\Delta z =
0.33$ convolved with a Gaussian of $\sigma(z) = 0.2$.  Solid:  $\epsilon
= 0.8$;  dotted:  $\epsilon = 0$; dash-dot:  Baugh et al. (1999)
semi-analytic model for $R<23.5$ and no dust; dash-triple dot: $\epsilon
= -1.2$.
\label{fig:acfvsz}}

\figcaption{
Random errors in $A_w$ obtained from delete-$d$ jackknife.  Each panel
is for a given $R$ magnitude and redshift range, as indicated by the
legends, and each point is a particular subsample or extrapolated value.
Abscissa:  number of galaxies in subsample.  Ordinate:  standard
deviation in $A_w$ among 50 or 100 subsamples of the given size.
Jackknife results are given by solid circles, and open circles show the
extrapolation to the full galaxy sample.  Triangles show upper and lower
limits on error assuming that the various angular ($\theta$) bins in the
DD correlation are perfectly correlated (top triangle) and perfectly
independent (bottom triangle), respectively.
\label{fig:jack}}

\figcaption{
Masks for $R$-band BTC image used in correlation computations and
systematic error assessment.  Black areas are excluded, and white are
included.  The SNR (or total exposure) over the frame is divided for
this purpose into three ranges, low, medium, and high (cf. Figure
\ref{fig:syserr}).  Top row, left: default mask, which excludes saturated
stars and blemishes; center: vertical stripe containing possible
low-level galaxy surface density enhancement excluded; right: only areas
of low SNR included.  Middle row, left: only areas of medium SNR
included; center:  only areas of high SNR included;  right:  low SNR
included, with extra masking around two bright stars.  Bottom row, left:
medium SNR included, with extra masking around two bright stars; center:
high SNR included, with additional masking around two bright stars;
right: areas of both low and high SNR included.
  \label{fig:masks}}

\figcaption{
Variation in correlation strength as a result of varying sensitivity
over the $R$-band image.  Abscissa:  mid-point of redshift bin;  actual
values are 0.167, 0.5, and 0.833, but symbols are shifted for clarity.
Ordinate:  correlation strength $A_w=w(1\arcdeg)$ for subsample masked
as shown in Figure \ref{fig:masks}.  Legend describes masks and gives
proportion of sources admitted by each.  The lowest redshift bin shows
the greatest variation apart from outliers.  Estimator $w_\mathrm{LS}$
used with $2.5\times 10^5$ random points in each DR run, and 1000
iterations of 1000 points in each RR run; integral constraint computed
for each mask with 1000 iterations of 1000 random points.
\label{fig:syserr}}

\clearpage

\begin{figure}[h]
\resizebox{!}{5in}{\includegraphics{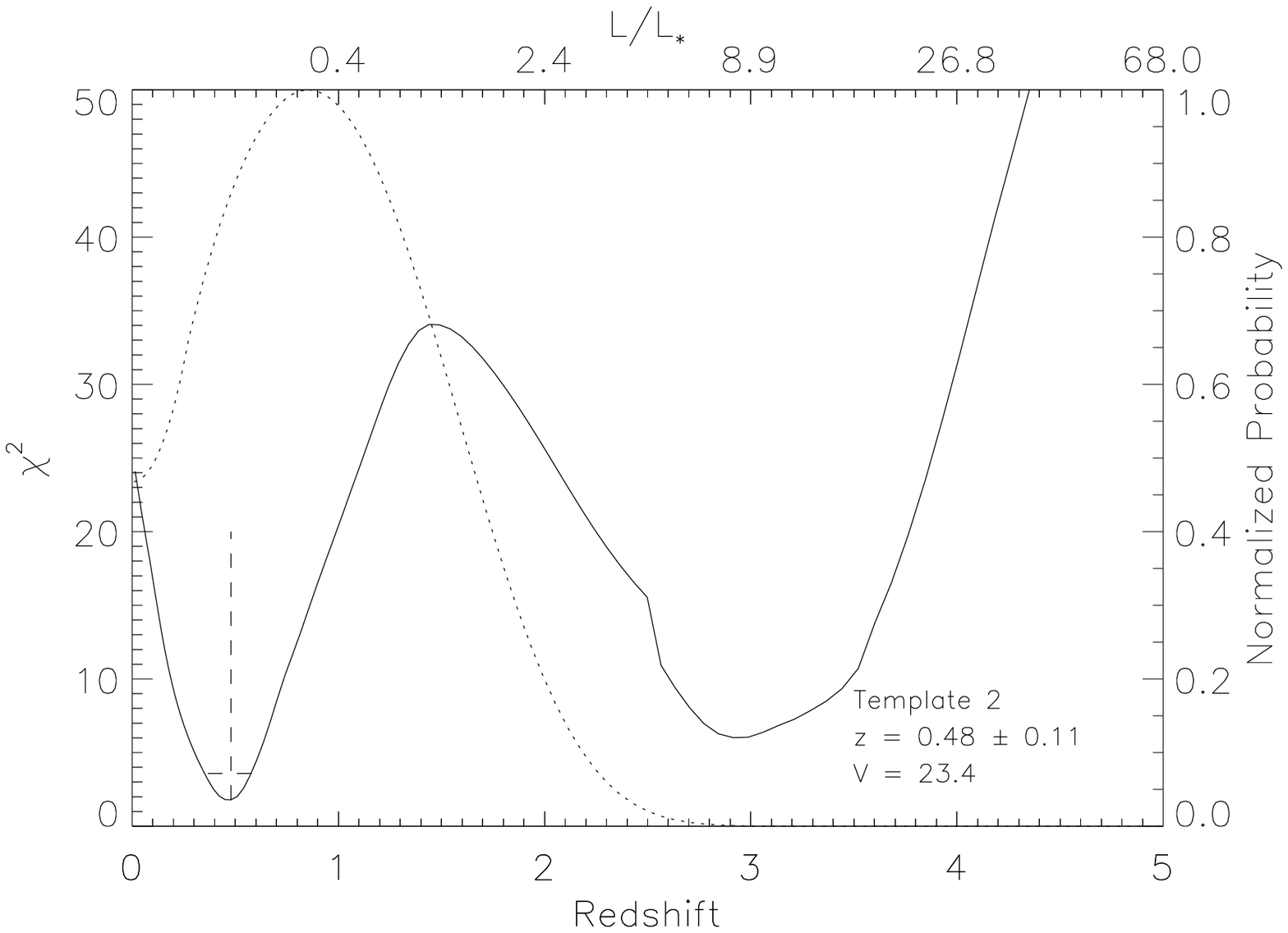}}
\end{figure}

\clearpage

\begin{figure}[h]
\resizebox{!}{6in}{\includegraphics{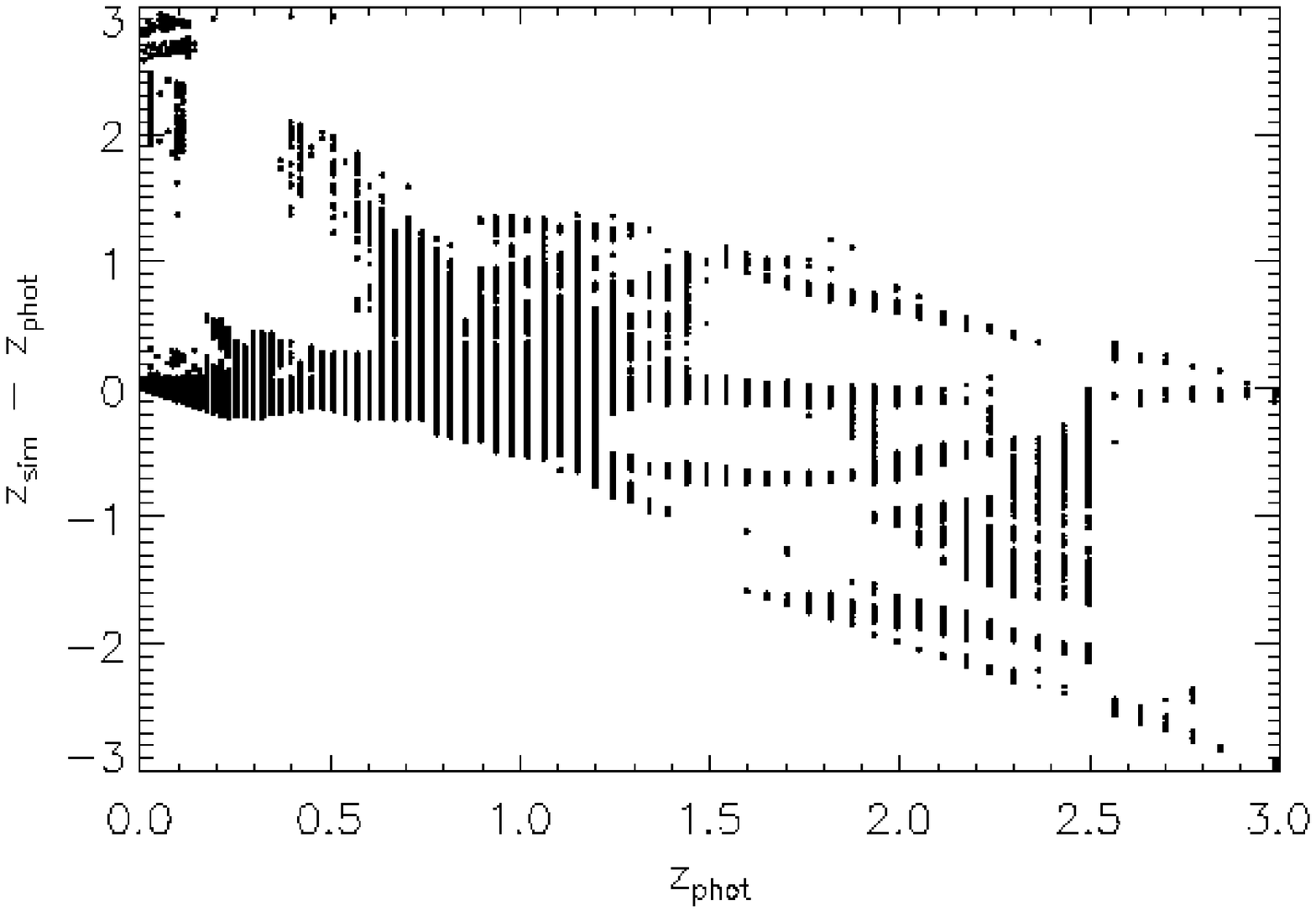}}
\end{figure}

\clearpage

\begin{figure}[h]
\resizebox{!}{6in}{\includegraphics{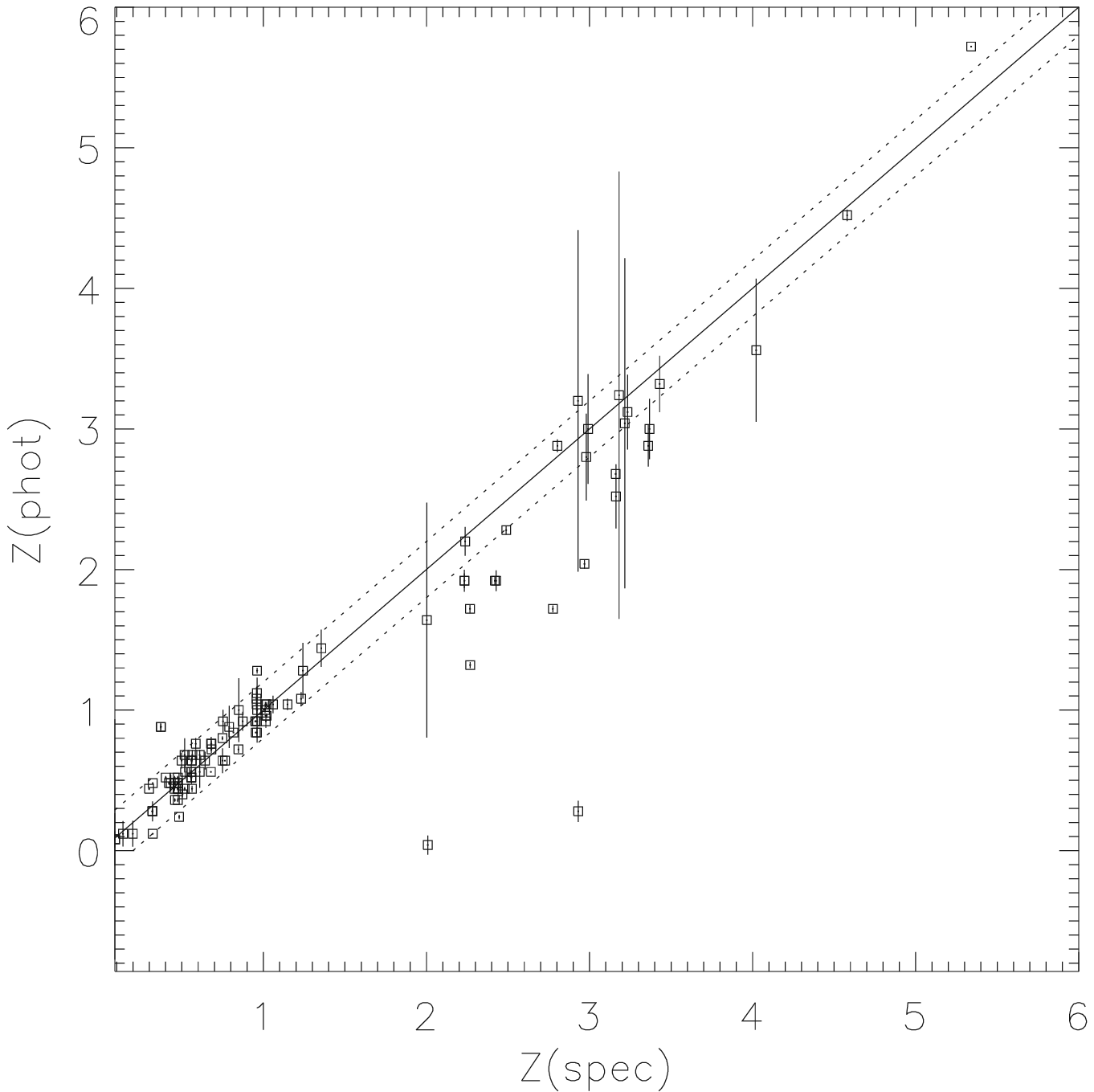}}
\end{figure}

\clearpage

\begin{figure}[h]
\hspace{-1in}
\resizebox{!}{6in}{\includegraphics{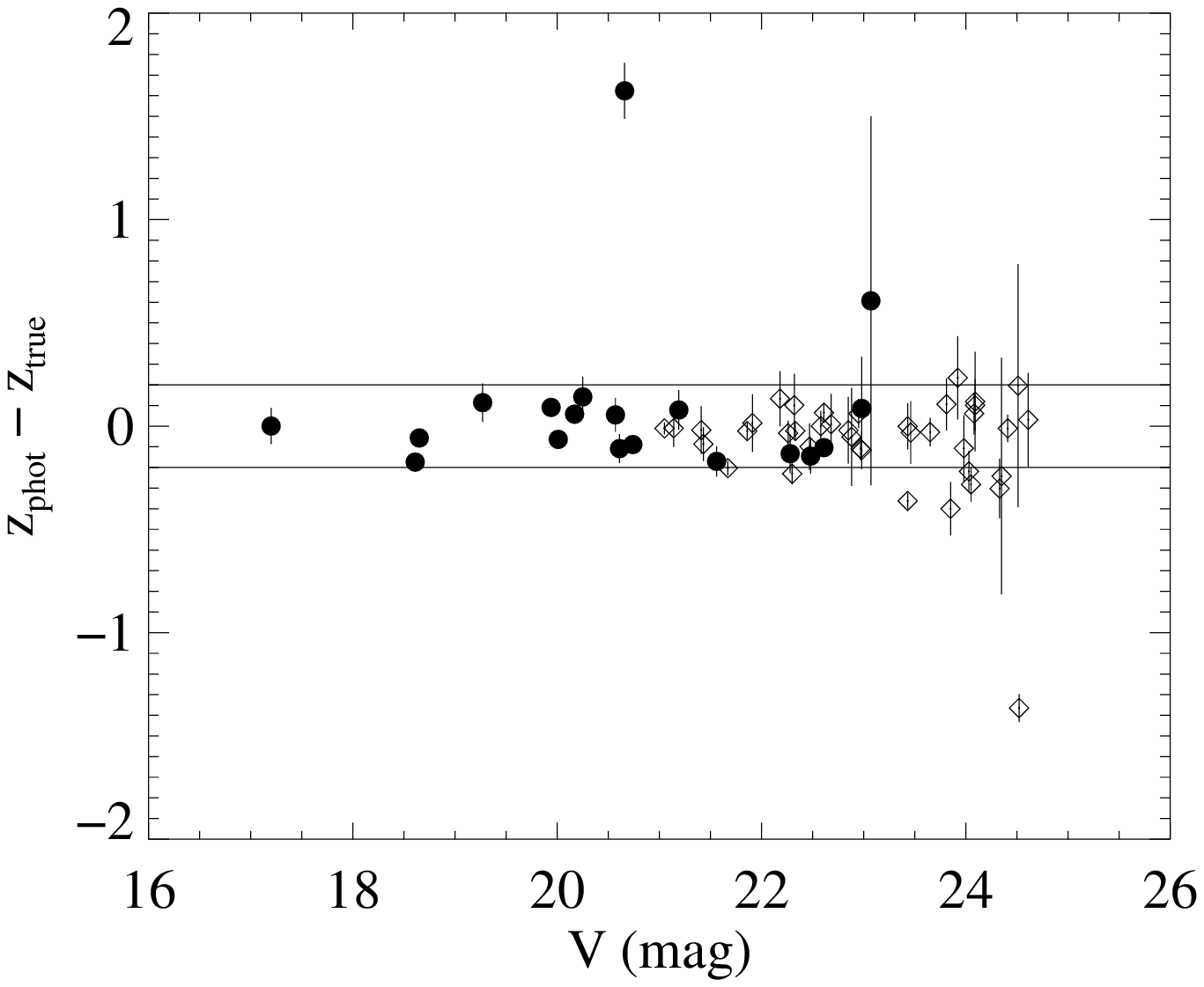}}
\end{figure}

\clearpage

\begin{figure}[h]
\resizebox{!}{7in}{\includegraphics{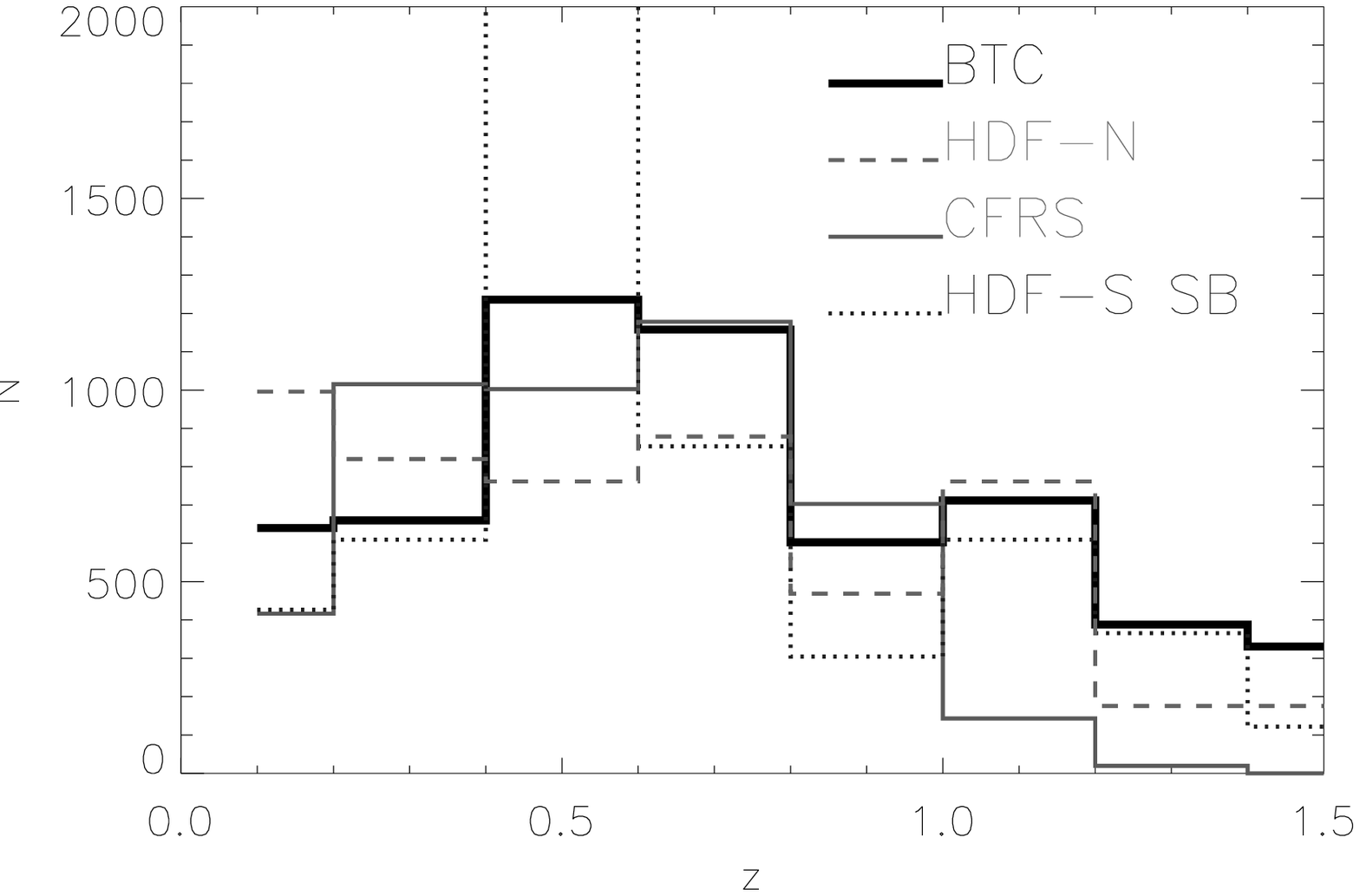}}
\end{figure}

\begin{figure}
\resizebox{\textwidth}{!}{\includegraphics{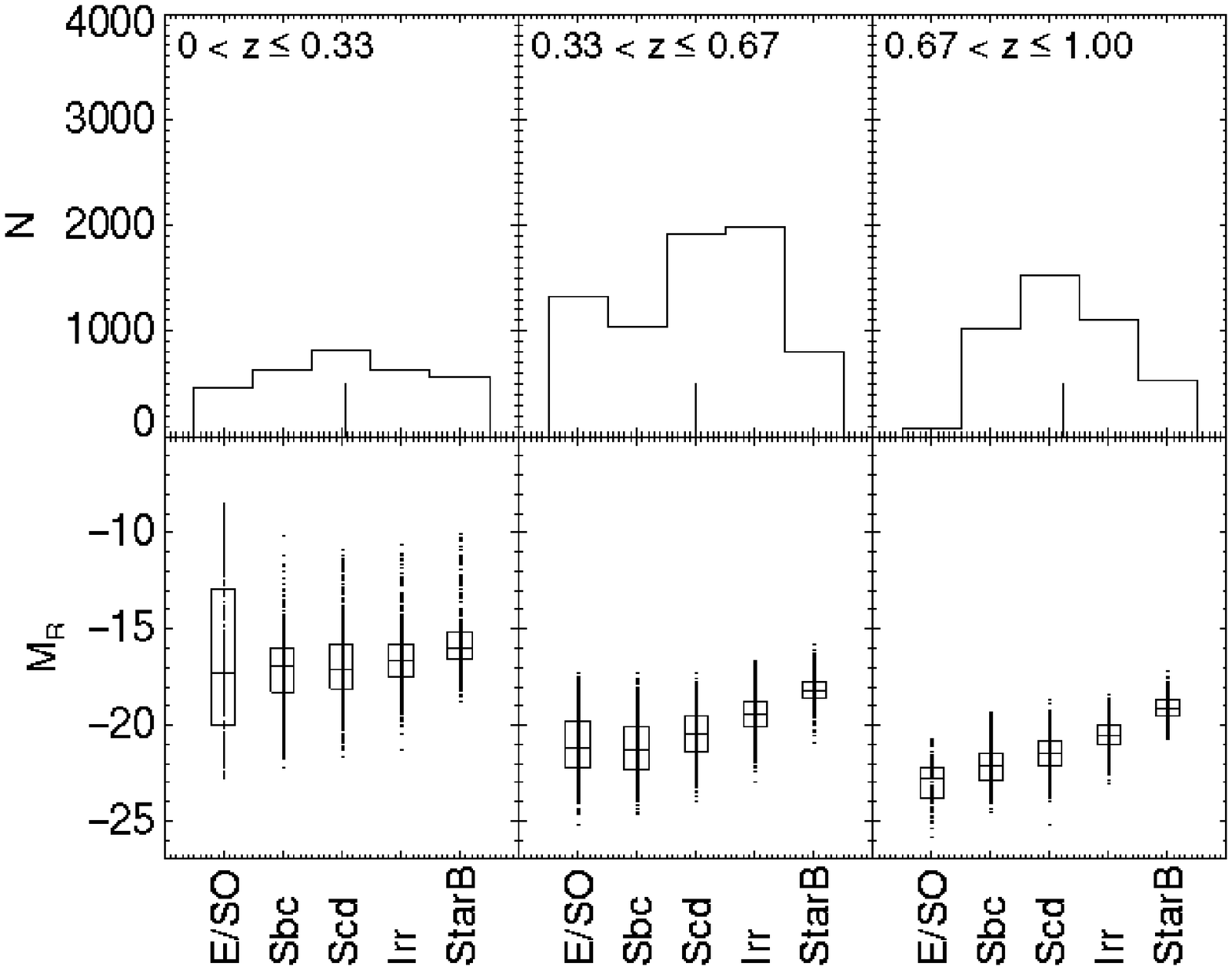}}
\end{figure}

\clearpage

\begin{figure}
\includegraphics{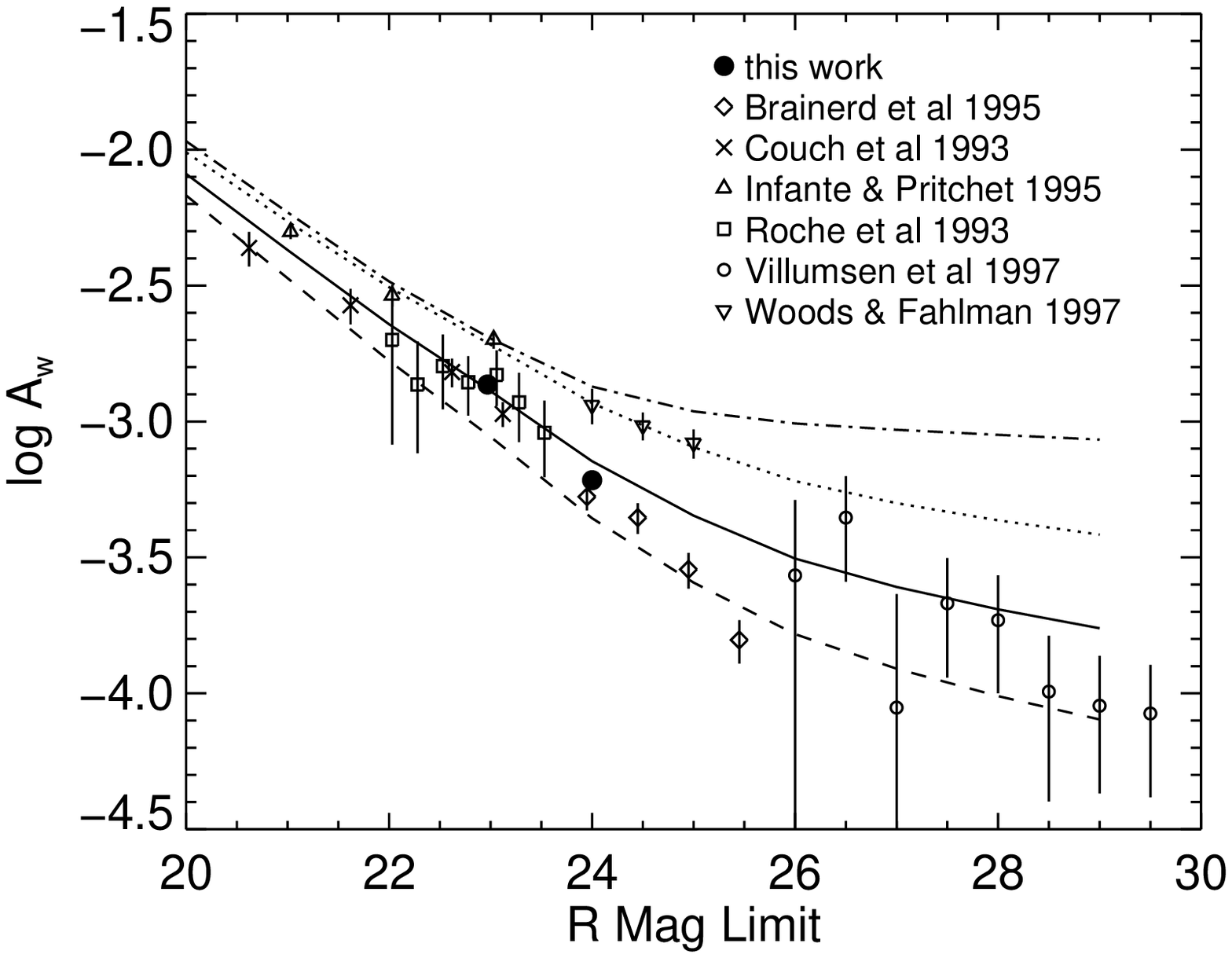}
\end{figure}

\clearpage

\begin{figure}
\includegraphics{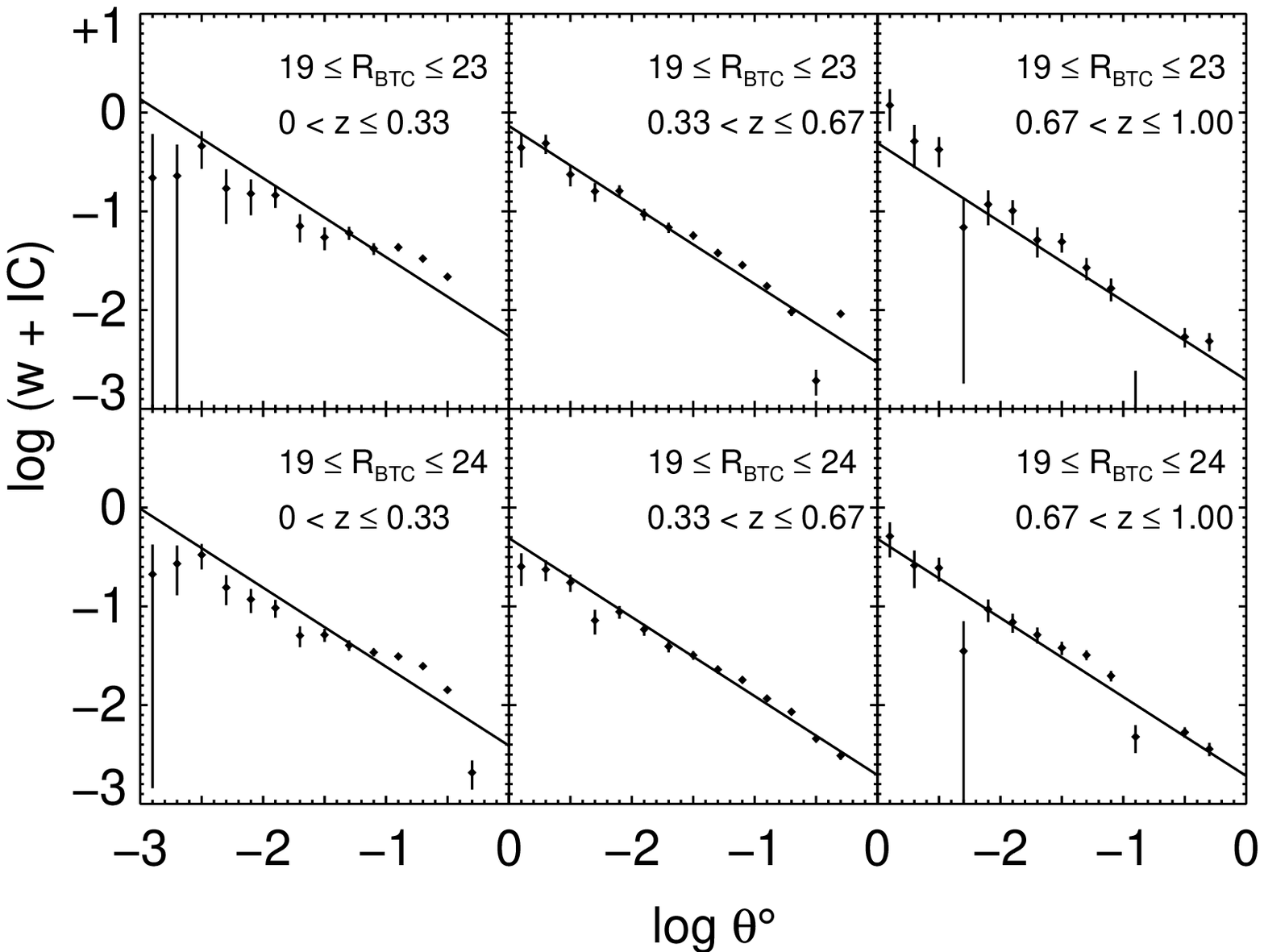}
\end{figure}

\clearpage

\begin{figure}
\hskip -1in
\resizebox{!}{6in}{\includegraphics{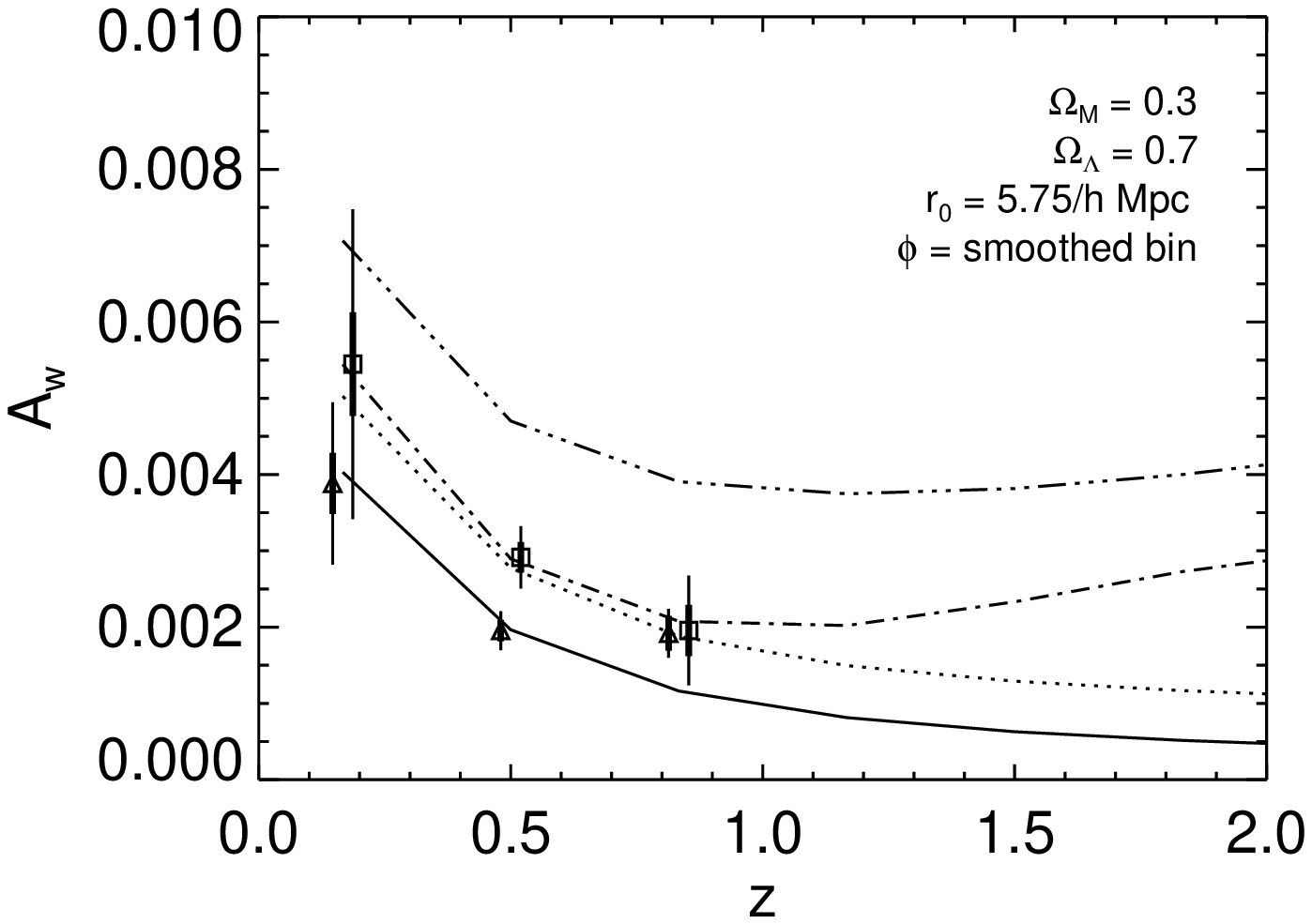}}
\end{figure}

\clearpage

\begin{figure}
\includegraphics{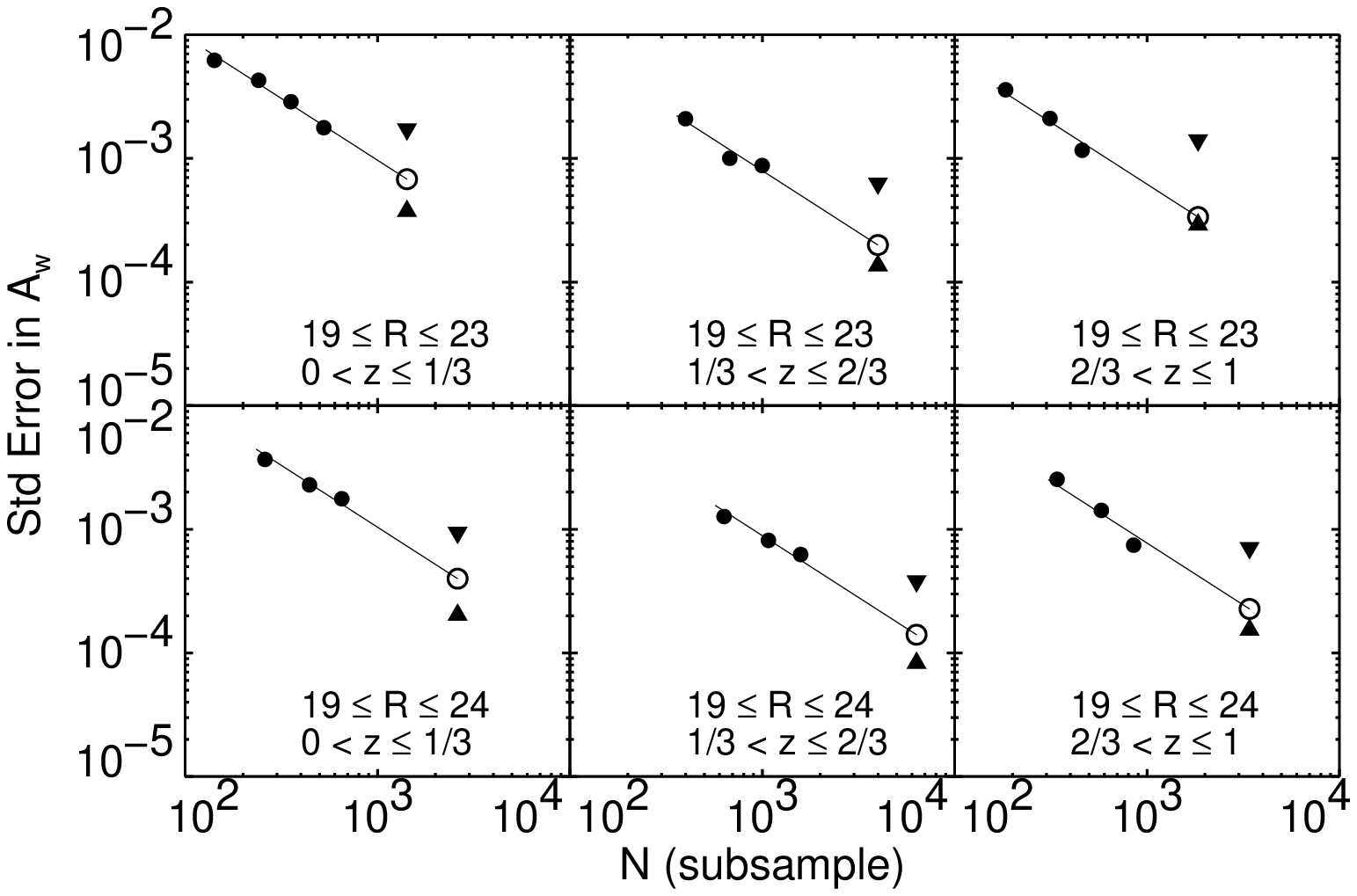}
\end{figure}

\clearpage

\begin{figure}
\resizebox{!}{5in}{\includegraphics{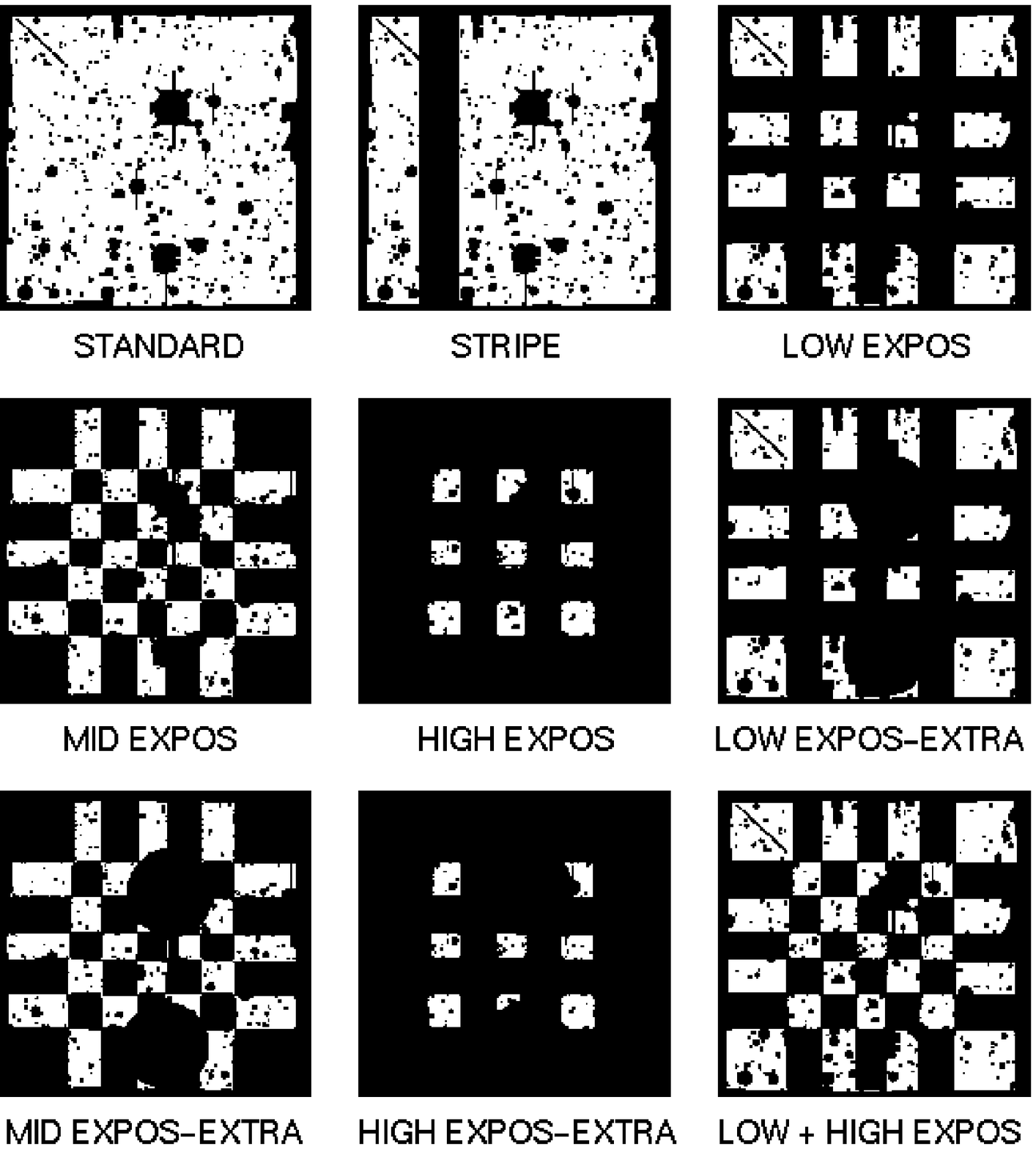}}
\end{figure}

\clearpage

\begin{figure}
\includegraphics{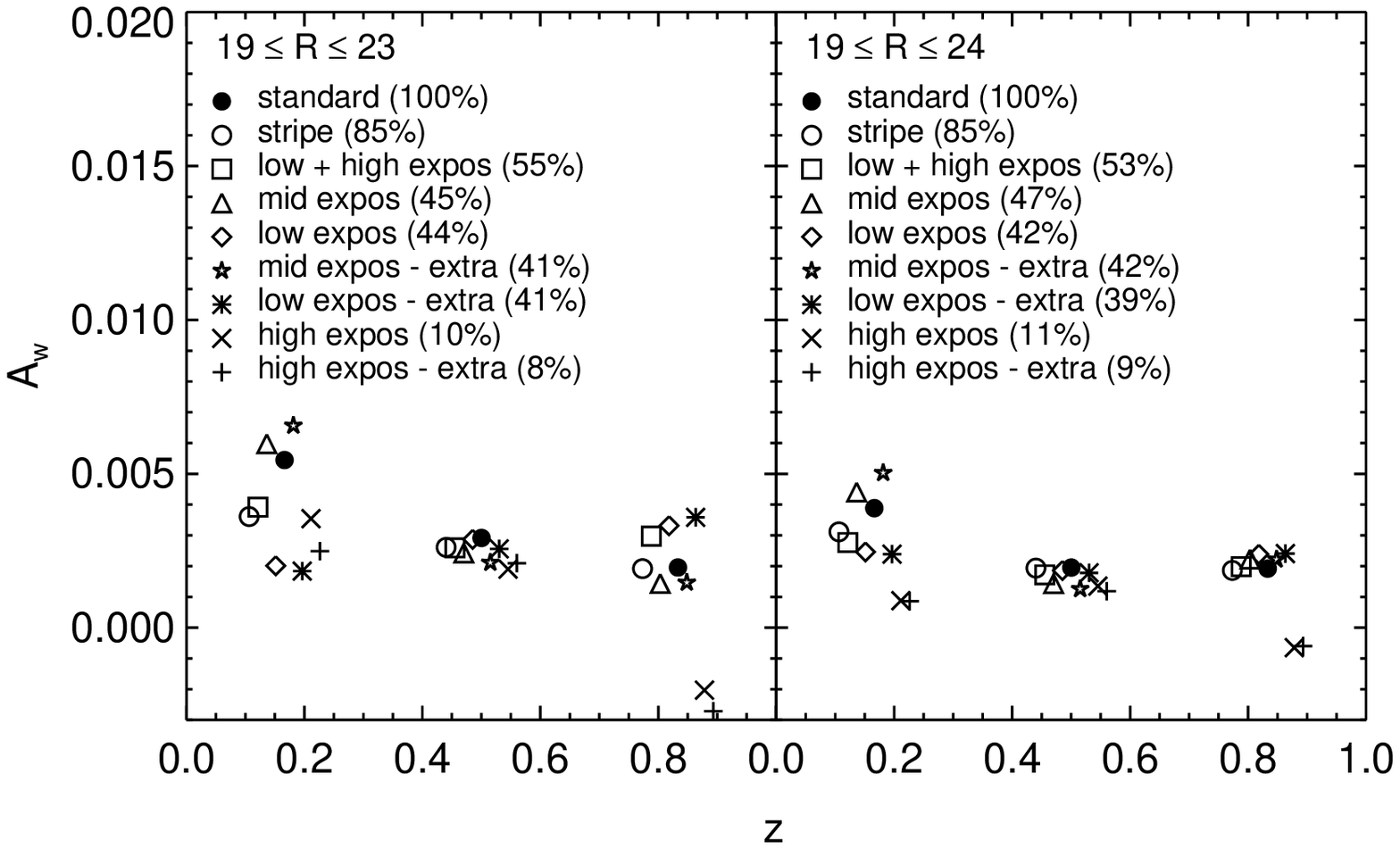}
\end{figure}

\clearpage

\end{document}